\documentclass{aa}  

\usepackage{graphics,psfig,rotate}

\begin{document}

\title{Long-Range Correlations in Self-Gravitating 
$N$-Body Systems}

\author{Daniel Huber and Daniel Pfenniger}

\institute{Geneva Observatory, CH-1290 Sauverny, Switzerland}

\date{Received July 2001 / Accepted January 2002}
\offprints{Daniel Huber\\ e-mail: Daniel.Huber@obs.unige.ch}

\authorrunning{D. Huber and D. Pfenniger}
\titlerunning{Correlations in Self-Gravitating Systems}

\abstract{Observed self-gravitating systems reveal often fragmented
  non-equilibrium structures that feature characteristic long-range
  correlations. However, models accounting for non-linear structure
  growth are not always consistent with observations and a better
  understanding of self-gravitating $N$-body systems appears
  necessary. Because unstable gravitating systems are sensitive to
  non-gravitational perturbations we study the effect of different
  dissipative factors as well as different small and large scale
  boundary conditions on idealized $N$-body systems. We find, in the
  interval of negative specific heat, equilibrium properties differing
  from theoretical predictions made for gravo-thermal systems,
  substantiating the importance of microscopic physics and the lack of
  consistent theoretical tools to describe self-gravitating gas.
  Also, in the interval of negative specific heat, yet outside of
  equilibrium, unforced systems fragment and establish transient
  long-range correlations. The strength of these correlations
    depends on the degree of granularity, suggesting to make the
    resolution of mass and force coherent.  Finally, persistent
  correlations appear in model systems subject to an energy flow.
   } \maketitle
\section{Introduction}

Most astrophysical structures result of gravitational
instabilities, from large scale cosmological structures down to
planets.  Yet, among the least understood topics in astrophysics 
we find galaxy formation and star formation, which both involve 
fragmentation and the nonlinear growth of structures occurring 
during the non-linear phases of gravitational instability.

Perhaps, one of the fundamental reason why fragmentation and structure
formation via gravitational instability appears so difficult is that
we lack of consistent theoretical tools allowing to combine gravity
with gas physics. Indeed, too often ignored is that classical
thermodynamics does not hold for gravitating systems, because these
are non-extensive in the thermodynamical sense (Landsberg
\cite{Landsberg72},\cite{Landsberg84}; Tsallis \cite{Tsallis99};
Plastino \& Plastino \cite{Plastino99}).  Actually, many other natural
systems do not respect the requisites of thermodynamics. Such systems
often feature interesting phenomena such as growing long range
correlations or phase transitions.  Among the symptoms of a
fundamental deep problem in gravitating systems is the appearance of
negative specific heat (Lynden-Bell \& Lynden-Bell \cite{Lynden77};
Lynden-Bell \cite{Lynden98}), which was seen for a long time as a
paradox in statistical mechanics, since negative specific heat was
thought to be impossible.

Presently, the only available approach to follow the nonlinear phases
of gravitational instabilities is to carry out numerical simulations.
Among all the existing methods, $N$-body techniques are thought to be
the most effective to simulate self-gravitating systems as well the
continuous case as the granular phases.
   
Yet, despite the considerable success of these methods in reproducing
many observed features, many fundamental problems remain.  As
mentioned above, the fragmentation and structure formation is not
clearly understood.  Related to this, CDM simulations conflict with
observations at galactic scales (Moore \cite{Moore99}, James et al.
\cite{Bullock01}, Bolatto et al. \cite{Bolatto02}), and no theory of the
ISM is presently able to {\it predict\/} the conditions of star
formation.  Most of the time the star formation process relies on
recipes with little physical constraint.
 
In situations were $N$-body simulations have success (e.g. hot stellar
systems) gravitational dynamics is sufficient to account for their
main global properties, additional microscopic physics can be neglected.
But when gravitational instability via fragmentation involves small
scale physics, the outcome may be strongly dependent on the properties
of the small scale physics.  In other words, in situations where 
the growth on singularities triggered by gravity is allowed, 
the chaotic nature of gravitating systems make them sensitive to 
the perturbations induced by non-gravitational physics.

Therefore it is important to understand the properties of $N$-body
systems subjected to various perturbations. For these purposes, 
a numerical study of perturbed, self-gravitating $N$-body systems
is carried out.

Among the relevant perturbations we expect that boundary conditions at
small and large scales, as well as dissipative factors can play a key
role.  In order to characterize the individual effects of
perturbations, in the tradition of analytical models, one is advised
to deliberately use simplified models.
 
A study of dissipative systems is important because such systems may
develop long-range correlations. In the typical ISM, radiative cooling
is very effective and induces a temporary energy-flow leading the
system far from equilibrium (Dyson \& Williams \cite{Dyson97}).  From
laboratory experiments it is well known that systems outside of
equilibrium may spontaneously develop spatio-temporal structures
(Glansdorff \& Prigogine \cite{Glansdorff71}; Nicolis \& Prigogine
\cite{Nicolis77}; Prigogine \cite{Prigogine80}; Melo \cite{Melo94}).

A permanent energy-flow is induced when energy loss due to dissipation
is replenished, that is, when the system is continuously driven, e.g., 
by time-dependent boundary conditions. Such systems may develop
persistent long-range correlations.  Astrophysical examples for this
are, the growth of structures in cosmological simulations, or the long
term persistence of filamentary structures in shearing flows (Toomre
\& Kalnajs \cite{Toomre91}; Huber \& Pfenniger \cite{Huber01a}; Wisdom
\& Tremaine \cite{Wisdom88}; Salo \cite{Salo95}; Pfenniger
\cite{Pfenniger98}). Among other things, the effect of time-dependent
potential perturbations on dissipative self-gravitating spheres
is studied in this paper.

In the next Section we briefly review some theoretical results of
the thermodynamics of self-gravitating isothermal spheres. The model is
presented in Sect. 3 and the applied methods to carry out the
structure analysis for the simulated systems are explained in Sect. 4.
The results are presented in Sect. 5-7.  In Sect. 5 quasi-equilibrium
states of $N$-body models are compared with analytical predictions.
Sect. 6 is dedicated to a study of long-range correlations appearing
in the interval of negative specific heat during the collapsing
transition of gravitating systems. Finally, in Sect. 7 the evolution 
of systems subjected to an energy-flow is discussed.

\section{Gravo-thermal Statistics}

By confining a gravitational system within a sphere, we can compare
our numerical work with previous analytical studies using the tools of
statistical mechanics and point out some discrepancies. 
Although the theory, sometimes referred to as
gravo-thermal statistics, is not fully consistent, since the
applied Gibbs-Boltzmann entropy is derived  under the assumption of
extensivity (Taruya \& Sakagami \cite{Taruya01}), it yields some
important and instructive results.

Before we present the model and the results let us here
briefly review some theoretical findings (see Padmanabhan
\cite{Padmanabhan90} for an extended discussion on the topic).
 
Antonov (\cite{Antonov62}) and Lynden-Bell \& Wood (\cite{Lynden68})
studied theoretically the thermodynamics of self-gravitating
isothermal spheres and found anomalous behavior when compared with
classical thermodynamics of extensive systems.  An example of such an
anomalous behavior is the so called gravo-thermal catastrophe that
occurs when an isothermal gas of energy $E(<0)$ and mass $M$ is
released within a spherical box of radius greater than Antonov's
radius, $r_A=0.335 G M^2/(-E)$.  The analytical model predicts
that for such systems there is no equilibrium to go and nothing can
stop the collapse of the central parts.

Antonov's and Lynden-Bell \& Wood's investigations based on point-like
particles. Since then several studies were carried out with modified
(i.e., non point-like) particle potentials.  Hertel \& Thirring
(\cite{Hertel71}) modified the gravitational interaction potential by
introducing short-distance repulsive forces due to quantum degeneracy
and Aronson \& Hansen (\cite{Aronson72}) investigated the behavior of
self-gravitating hard-spheres. Finally, Follana \& Laliena
(\cite{Follana00}) applied softened interaction potentials. They found
all qualitatively the same result: Unlike the model of Lynden-Bell \&
Wood, there is always an equilibrium state for finite size particles.
However, a phase transition, separating a high energy homogeneous
phase from a low energy collapsing phase with core-halo structure
occurs in an energy interval with negative
microcanonical specific heat,
\begin{equation}
c_V= \left(\frac{\partial\varepsilon}{\partial T}\right)_V=
-\beta^2\left(\frac{\partial\varepsilon}{\partial\beta}\right)_V<0\;,
\end{equation}
where $\varepsilon$ is the total energy and $\beta=1/T$ is the 
inverse temperature.  

There are fewer theoretical works done with the grand canonical
ensemble (e.g. de Vega, Sanchez \& Combes \cite{deVega96})  where mass would be
allowed to be exchanged with the environment, therefore we limit the
scope of this study to the canonical ensemble, that is energy can be
exchanged with the environment, but not the mass and angular momentum.

Note that an ensemble allowing to exchange also angular momentum would
be very relevant for astrophysical situations, but few theoretical
works have been made on this important aspect (Laliena
\cite{Laliena99}; Fliegans \& Gross \cite{Fliegans01}).
 
\section{Model}

A spatially isolated, spherical $N$-body system is studied.
The $N$ particles of the system are accelerated by gravity
and forces induced by boundary conditions and perturbations,
respectively. In all, a particle can be accelerated by 
five force types:
1.) confinement 2.) energy dissipation due to velocity dependent
friction 3.) external forcing 4.) long range attraction (self-gravity) and
5.) short range repulsion.  The five
force types are discussed in detail in Sects. \ref{seca}-\ref{sece}.

\subsection{Units}

A steep potential well confines the particles to a sphere of radius
$R\approx1$ (see Sect. \ref{seca}).  The total mass $M$ and the
gravitational constant $G$ are equal one as well, $G=M=1$. The
dimensionless energy is the energy measured in units of $GM^2/R$, that
is in our model $\approx1$. Thus energies and other quantities with
energy dimension (like the temperature) can be considered as
dimensionless.

\subsection{Confinement}
\label{seca}
 
In order to keep the model simple and to enable a comparison with
established theoretical results of canonical isothermal spheres we
apply a confinement that prevents gravitationally unbound particles
from escape and keeps thus the particle number constant.

The confinement is realized through a steep potential well
\begin{equation}
\Phi_{\rm conf} \propto R^{16}\;,
\end{equation}
where $R$ is the distance from the center of our spatially isolated
system. The walls of the potential well are steep, so that the
particles are only significantly accelerated by the confinement when
they approach $R=1$. However, the potential walls must not be too
steep, to avoid unphysical accelerations near the wall, due to a
discrete time-step.

Contrary, to non-spherical reflecting enclosures the applied potential well
conserves angular momentum.

\subsection{Energy Dissipation}
\label{dissip}
The effect of different dissipation schemes is studied:
Local dissipation, global dissipation and a scheme similar 
to dynamical friction. For convenience we call the latter hereafter 
``dynamical friction''. 

\subsubsection{Local Dissipation}
\label{loco}

The local dissipation scheme represents 
inelastic scattering of interstellar gas constituents.
That is, friction forces are added, that depend 
on the relative velocities and positions of the neighboring
particles. 

A particle is considered as a neighbor if its distance is
$r<\lambda\,\epsilon$, where $\epsilon$ is the softening length and
$\lambda$ is a free parameter. The friction force has the form:
\begin{equation}
F_i=\left\{\begin{array}{r@{\;\;:\;\;}l}
\Lambda\frac{m^2 r_i}{(r^2+\epsilon^2)^{3/2}}r^{2-\eta}
(\frac{\vec{V}\cdot\vec{r}}{r})^{\zeta} & r<\lambda\epsilon\;{\rm
and}\;\vec{V}\cdot\vec{r}<0\\
0 & {\rm otherwise},
\end{array}\right.
\end{equation}
where $V$ is the relative velocity of two neighboring particles,
$\zeta$, $\eta\ge 0$ are free parameters and $i=x,y,z$.  The term
$\vec{V}\cdot\vec{r}/r=V\cos\vartheta<0$ ensures a convergent flow and that
dissipation affects only the linear momentum.  As a consequence
angular momentum is locally conserved.

For $\epsilon\ll r < \lambda\,\epsilon$ the friction force is,
\begin{equation}
\label{lodi}
F_i\propto \frac{(V \cos\vartheta)^{\zeta}}{r^{\eta}}\, e_i\;,
\end{equation}
where $e_i$ is the $i$th component of the unity vector. Thus the
friction force increases with the relative velocity and decreases with
the distance, provided that $\zeta,\;\eta>0$. The condition 
$r<\lambda\,\epsilon$ ensures the local nature of the energy
dissipation. 

\subsubsection{Global Dissipation}

The global dissipation depends not on the relative velocity $V$, but on
the absolute particle velocity $v$. The global friction force is,
\begin{equation}
F_i=-\alpha\, v_i\;,
\end{equation}
where $\alpha$ is a free parameter.
The same friction force was already used in the shearing box
experiments of Toomre \& Kalnajs (\cite{Toomre91}) and 
Huber \& Pfenniger (\cite{Huber01a},\cite{Huber01b}).

\subsubsection{Dynamical Friction}

Setting the velocity-dispersion
$\sigma=1$, which is the dispersion of a virialized 
self-gravitating system with $M=R=1$,
Chandrasekhar dynamical friction formula 
can be parameterized in good approximation by, 
\begin{equation}
F_i=\Gamma\frac{v_i}{(4+v^3)}\;,
\end{equation}
where $\Gamma$ is a free parameter (Chandrasekhar
\cite{Chandrasekhar43}; Binney \& Tremaine \cite{Binney94}).  In
contrast to global dissipation this scheme has the feature that $F_i$
does not increase asymptotically with $v$. There is a maximum at
$v\approx 1.3$. Thus high speed particles in collapsed regions
dissipate no longer energy and gravitational runaway is impeded.

\subsection{Forcing Scheme}
\label{heating}

If the dissipated energy is replenished by a forcing scheme the system
can be subjected to a continuous energy-flow. Here the energy
injection is due to time-dependent boundary conditions, that is, due
to a perturbation potential. This perturbation should on the one hand
be non-periodic and quasi-stochastic, on the other hand it should
provide approximately a regular large-scale energy injection in time
average.  A simple perturbation potential, meeting these conditions,
has the linear form,
\begin{equation}
\Phi_{\rm pert}(x,y,z,t)=\gamma [B_x(t) x + B_y(t) y + B_z(t) z]\;,
\end{equation}
where $x, y$ and $z$ are Cartesian coordinates, $\gamma$ is a free
parameter determining the strength of the perturbation, and
\begin{eqnarray}
\label{eq}
B_x(t)&=&\sum^3_{i=1} A_{x,i}\sin(\omega_{x,i} t + \phi_{x,i})\nonumber\\
B_y(t)&=&\sum^3_{i=1} A_{y,i}\sin(\omega_{y,i} t + \phi_{y,i})\\
B_z(t)&=&\sum^3_{i=1} A_{z,i}\sin(\omega_{z,i} t + \phi_{z,i})\nonumber\;,
\end{eqnarray}
where $t$ is the time.  The amplitudes $A_{j,i}$ and the phases
$\phi_{j,i}$ are arbitrary fixed constants and remain unchanged for
all simulations $(j=x,y,z)$. The frequencies are given through
$\omega_{j,i}=\delta\Omega_{j,i}$, where $\Omega_{j,i}$ are arbitrary,
but not rationally dependent constants (to avoid resonances) as well.
$A_{j,i}$, $\phi_{j,i}$ and $\Omega_{j,i}$ lay down the form of the
potential. Then, the amplitude and the frequency of the perturbation
are controlled by two parameters, namely $\gamma$ and $\delta$,
respectively (Huber \cite{Huber01}).

The perturbations are a linear combination of stationary waves that
do not inject momentum.

If our system represents a molecular cloud then a perturbation similar
to those described above can be due to star clusters, clouds or other
massive objects passing irregularly in the vicinity. Indeed such
stochastic encounters must be quite frequent in galactic disks and we
assume that the average time between two encounters is,
\begin{equation}
 \tau_{\rm pert}=\frac{1}{f_{\rm pert}} \ga \tau_{\rm dyn}\;, 
\end{equation}
where $f_{\rm pert}$ is the mean frequency of the encounters and
$\tau_{\rm dyn}$ is the dynamical time.  Thus, these encounters can
provide a continuous low frequency energy injection on large
scales. 

Such a low frequency forcing scheme induces at first ordered particle
motions, which are transformed in the course of time to random thermal 
motion, due to gravitational particle interaction.

\subsection{Self-Gravity, Repulsion and Force Computation}
\label{sece}

\begin{figure}
\centerline{
\psfig{file=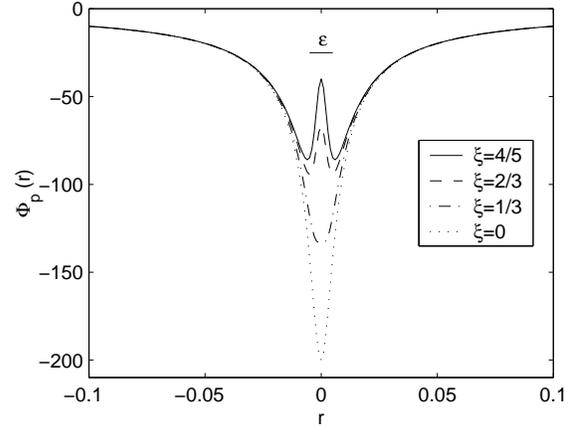,angle=90,width=7.5cm}}
\caption{\label{pot} The potential as a function of $r$ in units of
  $Gm/R$, where $m$ is the particle mass. A test
 particle at $|r|<\epsilon\sqrt(3\xi-1)$ feels a repulsive force if
 $\xi > 1/3$. The size of the softening parameter $\epsilon$ is
 indicated, as well.}
\end{figure}

Interaction forces include self-gravity and repulsive 
forces whose strength can be adjusted by a parameter. 
Repulsive forces may be a result of sub-resolution
processes such as star-formation or quantum degeneracy.
The particle interaction potential reads,
\begin{equation}
\label{equpot}
\Phi_p (r)=-\frac{Gm}{\sqrt{r^2+\epsilon^2}}\left
      (1-\xi\frac{\epsilon^2}{(r^2+\epsilon^2)}\right)\;,
\end{equation}
where $m$ is the particle mass and $r$ is here
the distance from the particle
center. On larger scales $(\ga\epsilon)$ the potential is a Plummer
potential $\Phi_{\rm Pl}$ 
(Plummer \cite{Plummer11}; Binney \& Tremaine \cite{Binney94}).
On small scales $(\la\epsilon)$  the deviation from a Plummer
potential and the strength of the repulsive force is determined 
by the parameter $\xi$. 
The interaction potential becomes for instance repulsive in the range
$|r|<\epsilon\sqrt(3\xi-1)$ if $\xi>1/3$ (see Fig.~\ref{pot}). 
If $\xi=0$, $\Phi_{\rm p}=\Phi_{\rm Pl}$ holds. 

The interaction forces are computed on the Gravitor Beowulf 
Cluster\footnote{http://obswww.unige.ch/\~{}pfennige/gravitor/gravitor.html} 
at the Geneva Observatory with a parallel
tree code. This code is based on the Fortran 
Barnes \& Hut (\cite{Barnes86}, \cite{Barnes89}) tree algorithm, 
and has been efficiently parallelized for 
a Beowulf cluster.  It is available on request.

The time-integration is the leap-frog algorithm with uniform
time-step, which ensures the conservation of the symplectic structure
of the conservative dynamics.  The time-step is,
\begin{equation}
\Delta t \le 0.1\frac{\epsilon}{\sigma_{\rm v}}\;,
\end{equation}
where $\sigma_{\rm v}$ is the velocity-dispersion of the initial state.

The accuracy of the force computation is given through the tolerance
parameter, which is for the studies presented in this paper,
$\theta\le 0.58$.

\subsection{Code Testing}

In order to test the code, the evolution of the
angular momentum is checked. We find that the angular momentum $J$
of a system with $10000$ particles and local dissipation scheme, 
which is initially $J=3.4\times 10^{-5}$ in units such that
energy is dimensionless, remains small, $J<4\times
10^{-5}$. This is illustrated by means of two examples. Two simulations with
equal dissipation strength are carried out. One with short range
repulsion, $\xi=2/3$, and the other without, $\xi=0$. After $10$
crossing times the angular momentum is $J=3.3\times 10^{-5}$ and
$J=3.8\times 10^{-5}$, respectively. The deviation from the initial
value is larger for the simulation without short range repulsion.
This is because the dissipation leads in this case to a
stronger mass concentration, i.e, to shorter particle distances.
Indeed, the potential energy after $10$ crossing times is $U\approx-12$
and $U\approx-17$ for the simulation with and without short range
repulsion, respectively.

\subsection{Parameters}

The different dissipation schemes, the forcing scheme and the
interaction potential with the short-distance repulsive force have
several free parameters. In order to do a reasonably sized parameter
study the particle number has to be limited to a maximum of
$N=160000$. However, important is not the absolute particle number, 
but the effect of changing it. Thus, also $N$ is varied.

The model parameters and their ranges are indicated in Table \ref{tab1}.

\begin{table*}
\begin{center}
\begin{tabular}{|c||r@{$,\ldots,$}l||l||l|} \hline
Parameter & 
\multicolumn{2}{c||}{Range} & 
Description & Effect \\ \hline \hline
%Description & \\ \hline \hline
$\delta$   & $1/\tau_{\rm dyn}$ & $2\pi/\tau_{\rm dyn}$ & 
Perturbation frequency & \\ 
$\gamma$   & $0$ & $0.7$& Perturbation factor & 
\raisebox{1.5ex}[-1.5ex]{Perturbation potential} \\ \hline %\hline

$\Lambda$    & $0$ & $5000$ & Local dissipation factor &\\
$\lambda$  & $2\,\epsilon$ & $50\,\epsilon$ & Dissipation volume&\\ 
$\eta$     & \multicolumn{2}{c||}{$0,\;1,\;2$} & Relative distance power& 
\raisebox{1.5ex}[-1.5ex]{Local dissipation}\\

$\zeta$    & \multicolumn{2}{c||}{$0,\;1,\;2$} & 
Relative velocity power&\\ \hline

$\alpha$   & $0$ & $50$ & Global dissipation factor&
Global dissipation\\ \hline

$\Gamma$   & $0$ & $30$ & Dynamical friction factor&
Dynamical firiction\\ \hline %\hline

$\epsilon$ & 
\multicolumn{2}{c||}{$0.0046,\;0.050$} & Softening length &\\

$\xi$      & \multicolumn{2}{c||}{$0,\;1/3,\;2/3,\;1$} & 
Repulsion strength &
\raisebox{1.5ex}[-1.5ex]{Interaction potential}\\ \hline 

$N$     & $6400$ & $160000$ & Particle number & Mass resolution\\ \hline 

\end{tabular} 
\caption{\label{tab1} Model paramaters, characterizing energy
  injection (perturbation potential), dissipation,
  interaction potential (self-gravity and repulsion) and mass resolution.}
\end{center}
\end{table*}

\section{Correlation Analysis}

In order to check if the perturbations of the gravitational system
induce characteristic correlations in phase-space the indices
$D$ and $\delta$ of the mass-size
relation and the velocity-dispersion-size relation
are determined.

The index of the mass-size relation can be found via,
\begin{equation}
D(r)=\left(\frac{{\rm d}\log M}{{\rm d}\log r}\right)(r)\;,
\end{equation}
where $M$ is the sum of the masses of all particles having
a relative distance $r$. The mass size relation is then
\begin{equation}
M(r)\propto r^{D(r)}\;, 
\end{equation}
where $r$ can be considered as the scale.  For a homogeneous mass
distribution and for a hierarchical fragmented, fractal structure the
index is a constant. For the latter case the index is not necessarily
an integer and lies in the interval, $D_{\rm T}<D<D_{\rm S}$, where
$D_{\rm T}$ and $D_{\rm S}$ are the topological dimension and the
dimension of the embedding space, respectively (Mandelbrot
\cite{Mandelbrot82}).

The index of the velocity-dispersion-size relation is 
determined by,
\begin{equation}
\delta(r)=\left(\frac{{\rm d}\log\sigma}{{\rm d}\log r}\right)(r)\;, 
\end{equation}
where $\sigma$ is the velocity-dispersion of
all particles having a relative distance $r$.
Observations of the interstellar medium
suggest a constant index $\delta=\delta_{\rm L}$ on
scales ${\cal O}(0.1)-{\cal O}(100)$ pc with
$0.3\la\delta_{\rm L} \la 0.5$. This is expressed by
Larson's law (e.g. Larson \cite{Larson81}, Scalo \cite{Scalo85}, 
Falgarone \& Perault \cite{Falgarone87}, Myers \& Goodman \cite{Myers88})
\begin{equation} 
\sigma\propto r^{\delta_L}\;.
\end{equation}

In order to preclude the effect of boundary conditions 
on the scaling relations,
upper and lower cutoffs have to be taken into account. 
The lower cutoff is given by the softening length.
An upper cutoff arises from the final system size.
Thus, the scope of application is for the 
velocity-dispersion-size and the mass-size
relation, $\epsilon<r<2R_{90}$, and
$\epsilon<r<R_{90}/2$, respectively, where
$R_{90}$ is the radius of the sphere centered at the 
origin which contains $90\%$ of the mass
(see Huber \& Pfenniger \cite{Huber01a}).

\section{Results: I. Properties of Equilibrium States in  
Numerical and Analytical Models}
\label{gravothermal}

Here, some $N$-body gravo-thermal experiments of systems with weak
dissipation, i.e., of systems in quasi-equilibrium are presented. The
results are compared with theoretical findings.

In order to cover a range of energy with the same experiment, a
convenient way is to introduce a weak global dissipation scheme
allowing to describe a range of quasi-equilibrium
states. Here weak means that $\tau_{\rm dis} \gg \tau_{\rm dyn}$,
where $\tau_{\rm dis}$ is the dissipation time-scale.  Then, the
results can be compared with theoretical equilibrium states.

\begin{figure}
\centerline{
\psfig{file=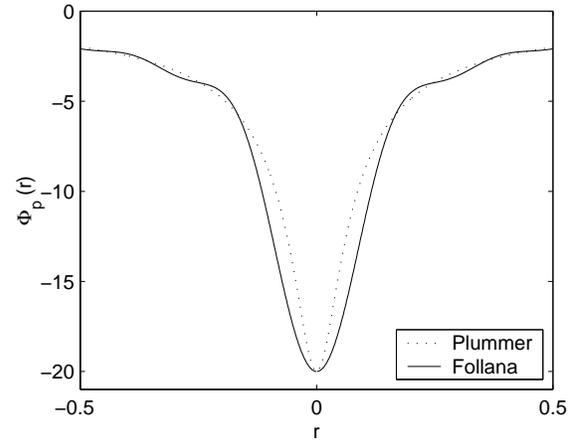,angle=90,width=7.5cm}}
\caption{\label{folpot} The Follana ($n=10$) and the Plummer
  ($\epsilon = 0.05$) potential, in units of $Gm^2/R$, where $m$ is the
  particle mass.}
\end{figure}

Follana \& Laliena (\cite{Follana00}) examined theoretically the
thermodynamics of self-gravitating systems with softened potentials.
They soften the Newtonian potential by keeping $n$ terms of an
expansion in spherical Bessel functions (hereafter such a regularized
potential is called a Follana potential).  This regularization allows
the calculation of the thermodynamical quantities of a
self-gravitating system.  The form of their potential is similar to a
Plummer potential with a corresponding softening
length. Fig.~\ref{folpot} shows a softened Follana
potential with $n=10$, and a Plummer potential with $\epsilon = 0.05$.

In their theoretical work Follana \& Laliena found for a mild enough
regularization ($n<30$) a phase transition below the critical energy
$\varepsilon_c\approx -0.335$ in a region with negative specific
heat. The transition separates a high energy homogeneous phase from a
low energy collapsed phase with core-halo structure.

We want to reproduce these findings by applying
a Plummer potential ($\xi=0$). Furthermore, 
the effect of a small-scale repulsive force
($\xi>1/3$) is studied.  

For these purposes, simulations with a weak global dissipation
strength, $\alpha=0.025$, are carried out.  The dissipation time is
then $\tau_{\rm dis} = 80\;\tau_{\rm dyn} = 56.6\;\tau_{\rm ff}$,
where the free fall time is, $\tau_{\rm ff}=\tau_{\rm dyn}/\sqrt{2}$.

Before we discuss the results, let us briefly present some model
properties. The initial state is a relaxed, unperturbed and confined
$N$-body sphere with total energy $\varepsilon=1$.

Assigning a particle a volume of $4\pi\epsilon^3/3$, the 
volume filling factor is, 
\begin{equation}
V_{\rm ff}=\frac{N \epsilon^3}{R^3}\;,
\label{vff}
\end{equation}
where $N$ is the particle number and $R$ is the radius of the system.
The volume filling factor is set $V_{\rm ff}=1$, meaning that the
force resolution is equal to the mass resolution. Then,
the particle number is, $N=8000$.
  
\begin{figure}
\centerline{
\psfig{file=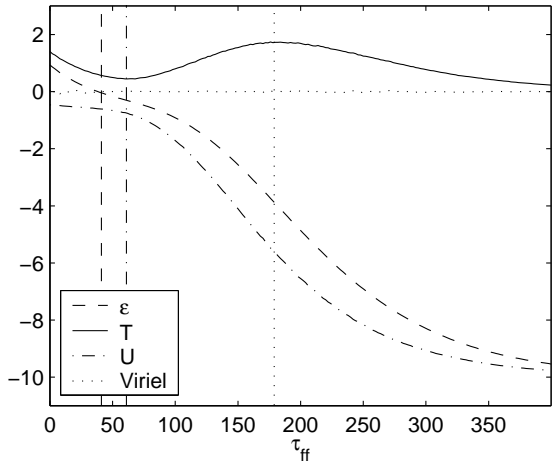,angle=90,width=7.5cm}}
\caption{\label{colenergy} The potential $U$, kinetic $T$ and 
  total energy $\varepsilon$ as a function 
  of time for a simulation 
  with global dissipation and Plummer softening ($\xi=0$). 
  The temporal evolution of the
  virial equation is indicated as well (${\rm Viriel} \equiv \ddot{I}/2$).
  The dashed vertical line depicts the time, when the system becomes
  fully self-gravitating. The dash-dotted and the dotted line mark the
  interval of negative specific heat.}
\end{figure}

The applied weak dissipation strength maintains the system
approximately thermalized and virialized (see Fig. \ref{colenergy}).
Here a system of $N$ particles is called virialized when the moment of
inertia $I$ is not accelerated,
\begin{equation}
\frac{\ddot{I}}{2}=2T+\sum_{i=1}^N F_i r_i\approx0\;,
\end{equation}
where $T$ is the kinetic energy, $r_i$ is the location of the $i$-th
particle and $F_i$ is the sum of all forces acting on the particle
(gravitation, friction, confinement).

\begin{figure*}
\centerline{
\psfig{file=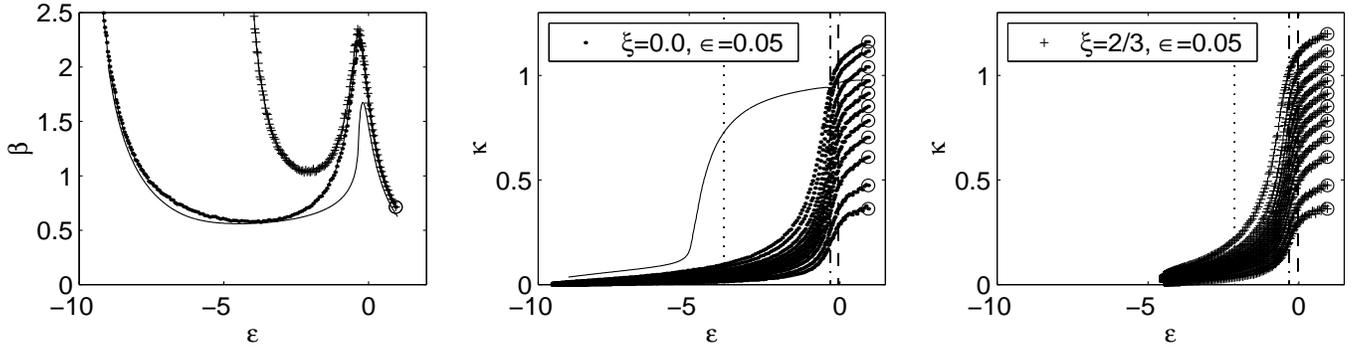,angle=90,width=\hsize}}
\caption{\label{colkappa1} 
Comparison of theoretical predictions and simulated
systems. The simulations are carried out with $N=8000$ and the
softening length is, $\epsilon=0.05$. Consequently, the volume filling
factor is, $V_{\rm ff}=1$. 
{\it Left}: Inverse temperature 
$\beta$ versus energy
$\varepsilon$ for models with softened potentials. 
The solid line indicates the theoretical result with
the Follana potential, $n=10$. The other curves depict the
evolution of the simulated systems. 
Crosses: repulsive potential
($\xi=2/3$). Dots: Plummer potential ($\xi=0$). 
The circle indicates the initial state of the simulations. 
The range of negative specific
heat corresponds to the range where the slope 
of $\beta(\varepsilon)$ is positive. 
{\it Middle}: The dotted lines describe the evolution of the 
Lagrangian radii $\kappa$
for the simulation with the Plummer softening 
potential ($\xi=0$).
Each curve depicts the radius of a sphere containing a 
certain mass fraction. The different mass fractions are: 
$\Delta M/M=\{5\%, 10\%, 20\%, \ldots ,80\%, 90\%, 95\%\}$. 
The solid line depicts the theoretical $95\%$-Lagrangian 
radius for the Follana potential. The dashed vertical 
line depict the moment when the system becomes fully
self-gravitating. The dash-dotted and the dotted line mark the
interval of negative specific heat.
{\it Right}: Idem for the simulation with the short range 
repulsive force.}  
\end{figure*}

The evolution of the inverse temperature
$\beta=1/T$ as a function of the total energy $\varepsilon$ for two 
simulations with $\xi=0$ and $\xi=2/3$, respectively, as well as the 
semi-analytical curve calculated by Follana \& Laliena are shown in
the left panel of Fig.~\ref{colkappa1}.

The interval of negative specific heat corresponds for the
simulation with the Plummer potential with theoretical
predictions. Also,
in accordance with predictions, a phase transition takes place, 
separating a high energy homogeneous phase, from a collapsed phase
in a interval with negative specific heat. Yet, the 
simulated phase transitions occur at higher energies.
To illustrate this,  
the evolution of the Lagrangian radii $\kappa$, are
shown in the middle and the right panel of Fig.~\ref{colkappa1}. 

In the high energy homogeneous phase the system is insensitive to the
short-distance form of the potential. Thus all systems enter the
interval of negative specific heat at the same energy.  However, the
collapsed phase is sensitive to the short-distance form. That is, the
system with repulsive short-distance force reenters the interval of
positive specific heat at a higher energy than those without such
forces (see left panel of Fig. \ref{colkappa1}).

Furthermore, the collapsed phase resulting from a simulation with a
Plummer potential is hotter and denser than those resulting from a
simulation with repulsive small scale forces. This can be seen in Fig.
\ref{colkappa1} as well.

For a smaller softening length theory expects a phase transition at higher
energies. Yet, already in simulations with a mild softening
$(\epsilon=0.05)$ the collapsing transition occurs straight after the
system has entered the interval of negative specific heat, which is
shortly after the system has become self-gravitating. Thus, the
collapsing transition can not occur at significantly 
higher energies for a smaller softening length.

\begin{figure*}
\centerline{
\psfig{file=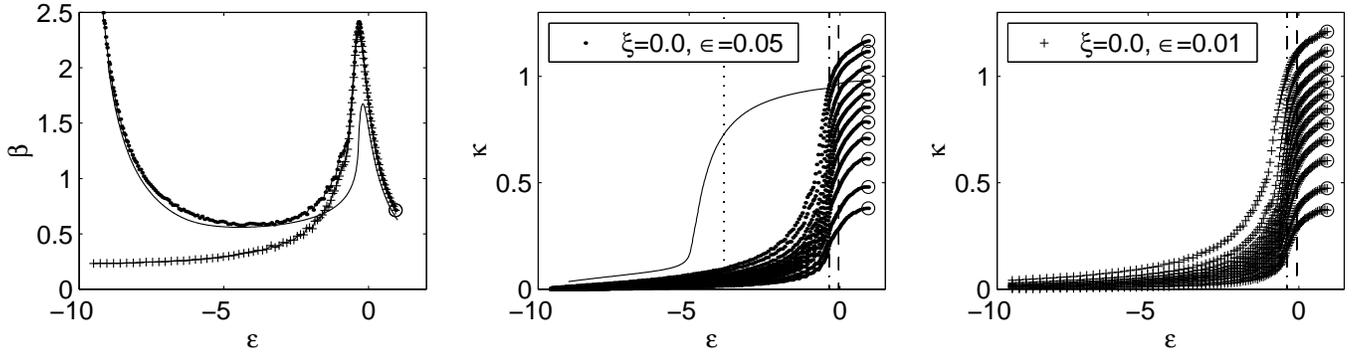,angle=90,width=\hsize}}
\caption{\label{colkappa2} 
Same as Fig. \ref{colkappa1} for two simulations with
$N=160000$ and with a Plummer potential $\xi=0.0$. 
Dots: $\epsilon=0.05$, $V_{\rm ff}=20.0$. 
Crosses: $\epsilon=0.01$, $V_{\rm ff}=0.16$.}
\end{figure*}

However, for systems with smaller softening length, the energy at which the
system reenters the zone of positive specific heat after the collapse
changes, i.e., it is shifted to smaller energies.  This is shown in
Fig. \ref{colkappa2}, where two simulations with $N=160000$ and
$\epsilon=0.05$ resp. $\epsilon=0.01$ are compared. 

Due to the higher particle number the dissipation time is in these
simulations reduced to, $\tau_{\rm dis}=4\tau_{\rm dyn}$. Thus,
dynamical equilibrium is not as perfect as previously. That is,
the acceleration of the
moment of inertia $I$ attains a maximum value of, $\ddot{I}/T=0.14$,
where $T$ is the kinetic energy. However, in general dynamical
equilibrium is well approximated and we expect, due to the experience
with simulations with varying particle number and dissipation time,
that the effect of the temporal deviation from equilibrium does not
affect qualitatively the results presented in Fig. \ref{colkappa2}.

The volume filling factor of the systems with $N=160000$ and
$\epsilon=0.05$ is, $V_{\rm ff}=20$, meaning that
the mass resolution is a factor $2.7$ larger than the force resolution.
Thus the system is less granular than in the previous simulations with
$N=8000$ and $V_{\rm ff}=1$ (see Fig. \ref{colkappa1}). Yet, this
does not affect the simulated quasi-equilibrium states and the
deviation from theoretical predictions remains.

\subsection{Discussion}

Some predictions made by analytical models, such as the interval of
negative specific heat, agree with findings resulting from $N$-body
models.  Yet, there appear also discrepancies, such as the way the
collapsing phase transition develops.

At first sight the deviating results are surprising.  Yet, the small scale
physics is different in the two models which may account for the
discrepancies.  Indeed, thermostatistics assumes a smooth density
distribution and is thus not able to account for two-body relaxation
effects in granular media.  However, a certain degree of granularity
is a inherent property of $N$-body systems. This may then lead to
discrepancies, especially in the unstable interval of negative
specific heat, where phase transitions may be sensitive to small scale
physics.

Whereas thermostatistics is too smooth to account for microscopic
physics in granular self-gravitating media such as the interstellar
gas, two-body relaxation is often too strong in $N$-body system due to
computational limitations. Especially in high force resolution
simulations where the force resolution is larger than the mass
resolution. We refer to this in Sect. \ref{corr} where we discuss,
among other things, the effect of the granularity on long-range
correlations appearing in the interval of negative specific heat.

A further discrepancy between nature, analytical models and $N$-body
models may be due to the entropy used in gravo-thermal statistics
(Taruya \& Sakagami \cite{Taruya01}).  Indeed, the entropy used in
analytical models to find equilibrium states via the maximum entropy
principle is the extensive Boltzmann-Gibbs entropy, that is in fact
not applicable for non-extensive self-gravitating systems.

Generalized thermostatistics including non-extensivity are currently
developed (Tsallis \cite{Tsallis88}; Sumiyoshi \cite{Sumiyoshi01}; Latora
et al. \cite{Latora01}; Leubner \cite{Leubner01}).  These formalisms
suggest that non-extensivity changes not all but some of the classical
thermodynamical results (Tsallis \cite{Tsallis99}; Boghosian
\cite{Boghosian99}), which agrees with our findings.

Because currently it is a priori not known which thermostatistical
properties change in non-extensive systems and consistent theoretical
tools are not available, analytical results must be considered with
caution.

\section{Results: II. Transient Long-Range Correlations in 
Gravitational Collapses}

\subsection{Effect of Dissipation}
\label{corr}
Properties of equilibrium states differing from theoretical
predictions in the interval of negative specific heat were found (see
above). Let us, now study the nonlinear structure growth in this
interval for an increased dissipation strength, i.e., outside of
equilibrium, when $\tau_{\rm dis}\la\tau_{\rm ff}$.

As previously, a relaxed, unperturbed, confined $N$-body sphere with
$\varepsilon=1$ serves as initial state for the simulations.

In order to dissipate the energy different dissipation schemes are
applied. Before we discuss the appearance of long-range correlations
in unstable dissipative systems let us briefly discuss the effect of
the different dissipation schemes on the global system structure.

The different applied dissipation schemes (see Sect. \ref{dissip})
lead to collapsed phases with different global structures. That is,
they have different mass fractions contained in the core and the halo.
A typical ordering is, ${\cal D}_{\rm global}>{\cal D}_{\rm
  dyn}>{\cal D}_{\rm local}$, where, ${\cal D}_{\rm global}$, ${\cal
  D}_{\rm dyn}$ and ${\cal D}_{\rm local}$ are the density contrasts
resulting from simulations with a global dissipation scheme, dynamical
friction and a local dissipation scheme, respectively.  The density
contrast is, ${\cal D}\equiv\log(\rho_{\rm cm}/\rho_0)$, where
$\rho_{\rm cm}$ and $\rho_0$ are the center of mass density and the
peripheric density, respectively. This means that for a global
dissipation scheme almost all the mass is concentrated in a dense core,
whereas a local dissipation scheme can form a persistent ``massive''
halo.

\begin{figure*}
\centerline{
\psfig{file=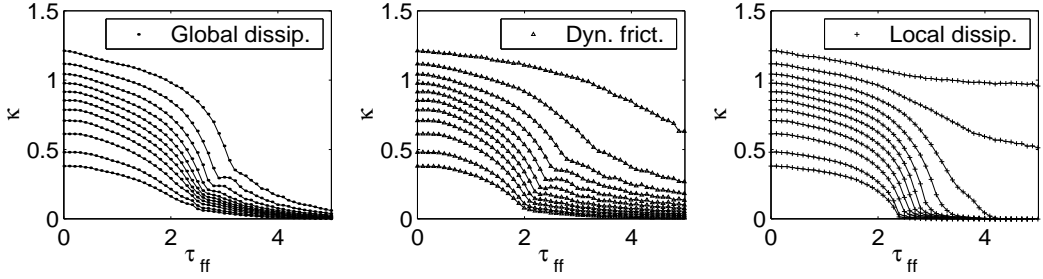,angle=90,width=14cm}}
\caption{\label{cooling} Evolution of the Lagrangian radii
for a system with global dissipation scheme, dynamical 
friction and a local dissipation scheme, respectively.
The Lagrangian radii depict, here and in the following figures,
spheres containing the following mass fractions: 
$\Delta M/M=\{5\%, 10\%, 20\%, \ldots ,80\%, 90\%, 98\%\}$.}
\end{figure*}

The evolution of the mass distribution resulting from simulations with
the different dissipation schemes is shown in Fig.  \ref{cooling}.
The collapse of the inner shells takes in all systems about the
same time. Yet, the uncollapsed matter distributed in the halo
is for the global dissipation scheme less than $2 \%$, whereas it
is $10 \%$ for the local dissipation scheme.  

Next, temporary long-range correlations that develop in unforced
gravitating systems are presented in dependence of the different
system parameters.

A sufficiently strong global dissipation, i.e., $\tau_{\rm
  dis}\la\tau_{\rm ff}$, of a gravitating system leads during the
nonlinear phase of the collapsing transition to fragmentation and
long-range phase-space correlations, so that the index, $\delta(r)$,
of the velocity-dispersion size relation, $\sigma\propto r^{\delta(r)}$,
becomes positive.

\begin{figure*}
\centerline{
\psfig{file=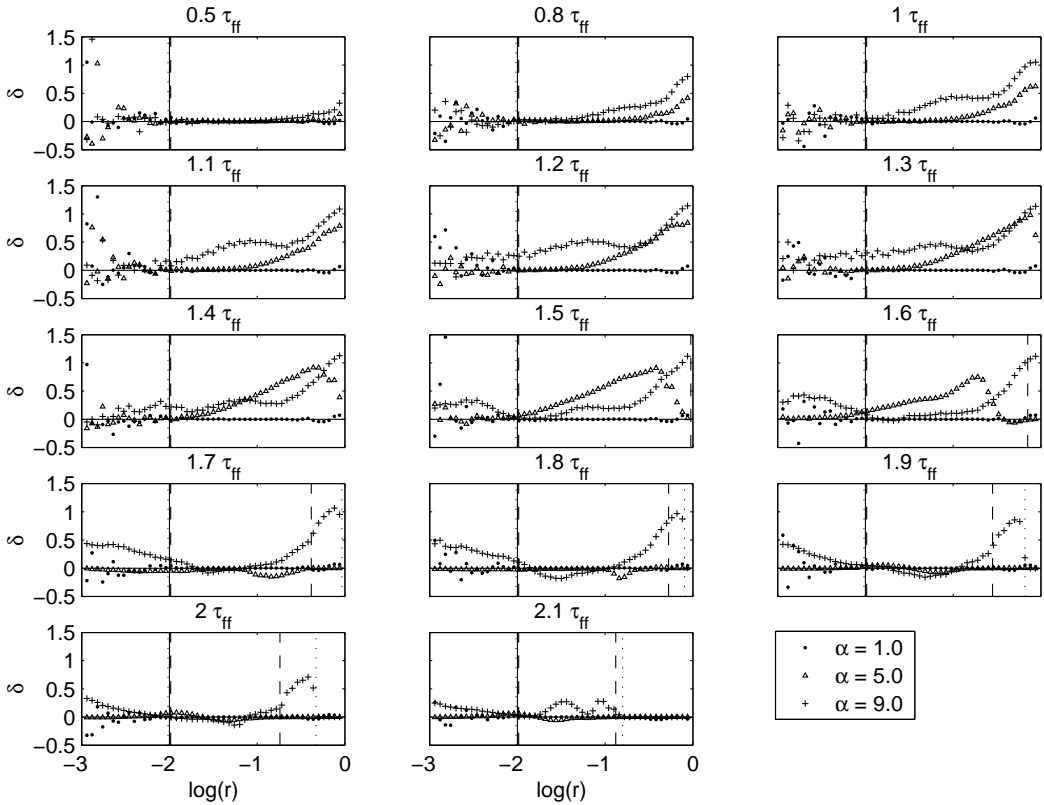,width=14cm}}
\caption{\label{kk3_1} Evolution of the index, $\delta(r)$, of the 
  velocity-dispersion-size relation, $\sigma\propto \delta(r)$, 
  during the collapsing transition
  for three simulations with {\bf global} dissipation scheme and different
  dissipation strength. The correlations result from simulations
  that were carried out with $160000$ particles. The solid,
  the dashed and the dotted vertical lines indicate the scope of
  application of the simulation with $\alpha=1.0$, $\alpha=5.0$ and
  $\alpha=9.0$, respectively. The lower cutoff is given by the
  softening length, that is here $\epsilon=0.01$, and the upper cutoff
  by $2R_{90}$.  The time is indicated above each panel.  The
  corresponding evolution of the Lagrangian radii and the specific
  heat are shown in Fig.  \ref{kk3kappa}.}
\end{figure*}

Fig. \ref{kk3_1} shows the evolution of the velocity-dispersion-size
relation, i.e., of $\delta$ for different dissipation strengths. The
relations result from 3 different simulations with global dissipation
schemes and different dissipation strengths. The dissipation strength is
given through the parameter $\alpha$. Here $\alpha=1.0,\;5.0,\;9.0$. 
This corresponds to $\tau_{\rm dis}=2.0,\;0.4\;,0.2\;\tau_{\rm ff}$.

Whereas the velocities remain uncorrelated in the simulation with
$\alpha=1.0$, $\delta$ becomes temporary positive over the whole dynamical
range in the simulations with the stronger dissipation.
  
The velocity correlations start 
to develop at largest scales after the system has become
self-gravitating. After the system has entered the interval of
negative specific heat the correlation growth is accelerated.
It attains a maximum and finally disappears when the collapse
ends. The dynamical range over which $\delta>0$ during 
maximum correlation is $\approx 2$ dex.  

Correlations at small scales straight above the softening
length are stronger for a stronger energy dissipation.

The maximum correlation established in the negative specific heat
interval persists for ${\cal O}(0.1)\;\tau_{\rm ff}$ and is characteristic for
the applied dissipation strength and softening. For instance, the
simulation with the strongest dissipation develops during $0.3\;
\tau_{\rm ff}$ a roughly constant $\delta$ over a range of 1.5 dex that
resembles Larson's relation.  Yet, correlations at small scales
decay rapidly and the index $\delta$ becomes even
negative at intermediate scales.

\begin{figure*}
\centerline{
\psfig{file=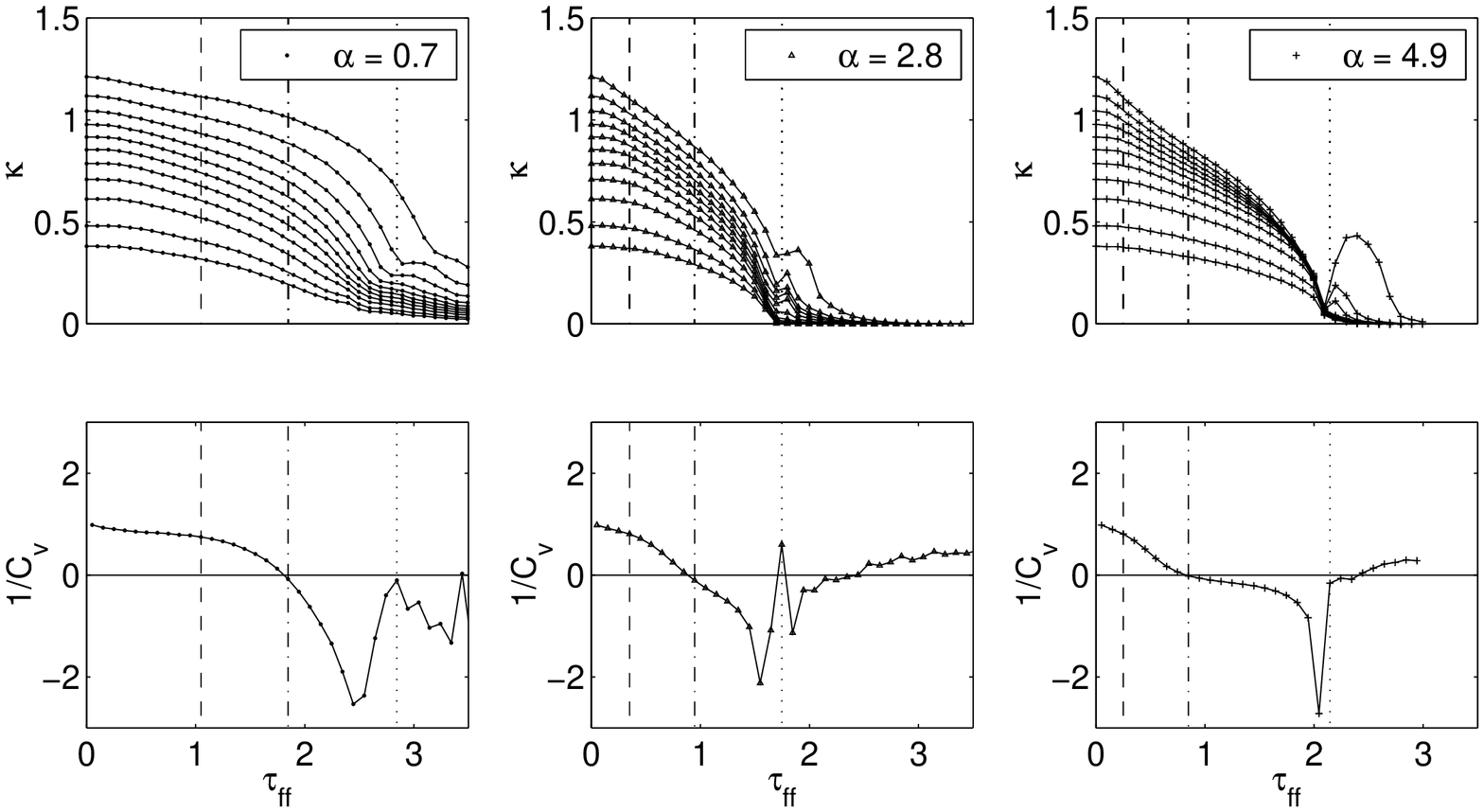,angle=90,width=14cm}}
\caption{\label{kk3kappa} Evolution of the Lagrangian radii (top)
  and the specific heat (bottom) 
  for three simulations with {\bf global} 
  dissipation and
  different dissipation strength. The dashed line marks the
  moment, when the system becomes fully self-gravitating. The
  dash-dotted line indicates the moment when the system 
  enters the
  zone of negative specific heat. The dotted line marks 
  the ``end'' of
  the collapse. The evolution of the corresponding phase-space
  correlations are shown in Fig. \ref{kk3_1}.}
\end{figure*}

The end of the collapse and with it the disappearance of the
correlated velocity structure is marked by a diverging
negative specific heat $c_V\rightarrow -\infty$. This is shown in Fig.
\ref{kk3kappa}.

Systems with dynamical friction and local dissipation develop also
velocity correlations. 

\begin{figure*}
\centerline{
\psfig{file=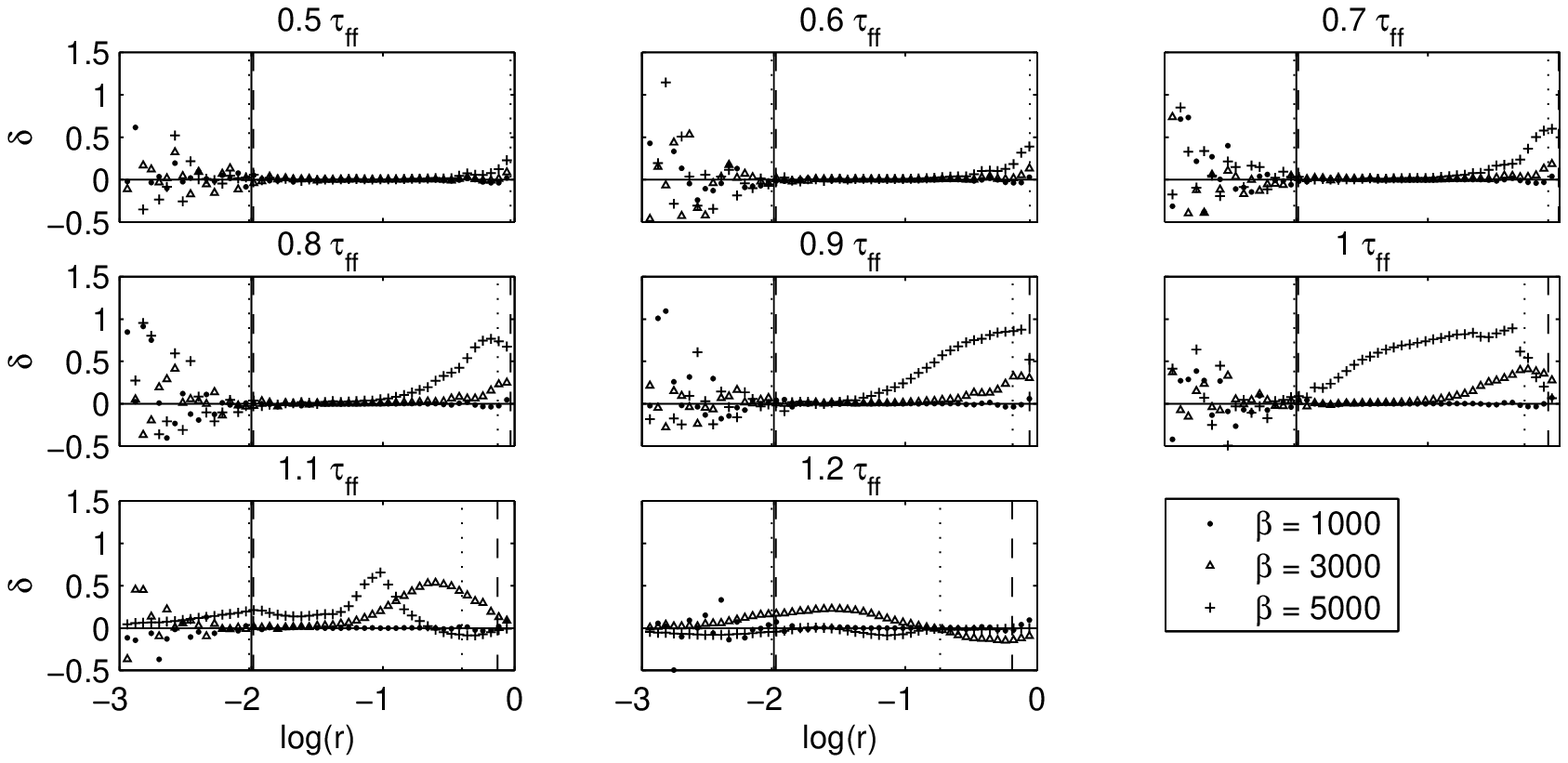,width=14cm}}
\caption{\label{kb_1} Evolution of the index of the 
  velocity-dispersion-size relation for three simulations with
  a {\bf local} dissipation scheme and different dissipation strengths.
  The particle number is, $N=160000$, and the softening length is,
  $\epsilon=0.01$.  The solid, the dashed and the dotted vertical
  lines indicate the scope of application of the simulation with
  $\beta=1000$, $\beta=2000$ and $\beta=5000$, respectively. The
  parameters determining the dependence of the local dissipation on
  the radius and the velocity are, $\eta=1$ and $\zeta=1$, respectively (see
  Sect \ref{loco}). The
  evolution of the corresponding Lagrangian radii and the interval of 
  negative specific heat are shown in Fig. \ref{kbkappa}.}
\end{figure*}

Fig. \ref{kb_1} shows the correlations resulting
from simulations with a local dissipation scheme. Contrary to the
global dissipation scheme, a strong local dissipation does not extend
the collapse of the whole system. Thus, local friction forces, that are
strong enough to develop correlations lead also to a fast collapse
and correlations persist accordingly a short time compared to
simulations with global dissipation. 

\begin{figure*}
\centerline{
\psfig{file=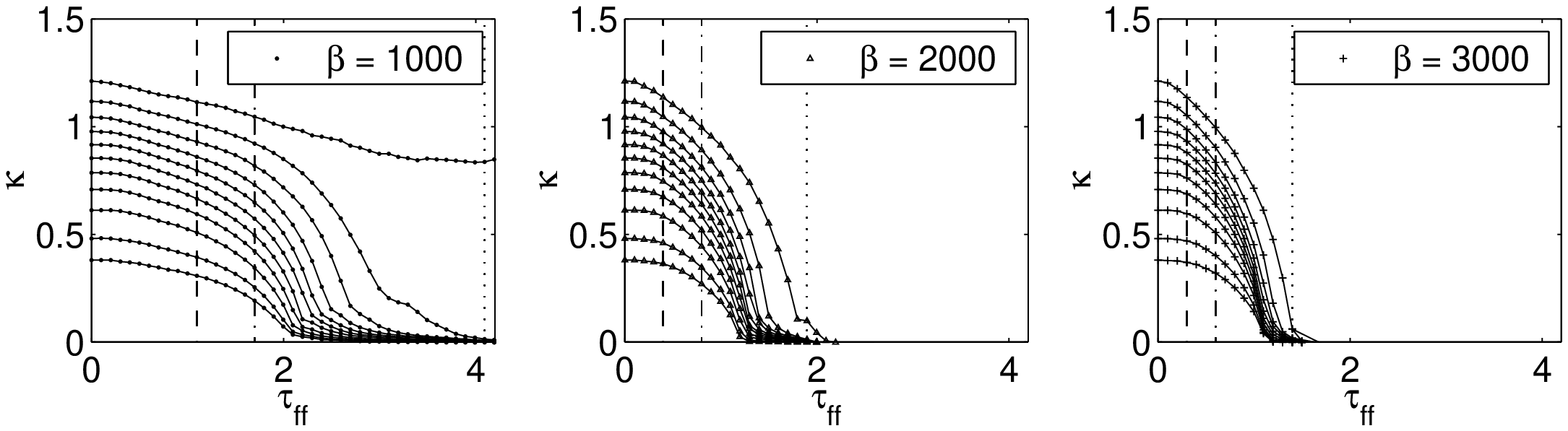,angle=90,width=14cm}}
\caption{\label{kbkappa} Evolution of the Lagrangian radii and
  the specific heat for three simulations with {\bf local} dissipation and
  different dissipation strength. The interval of negative specific
  heat is depicted by the dash-dotted and dotted line. The dashed line
  marks the moment when the whole system becomes self-gravitating. 
Fig. \ref{kb_1} shows the evolution of the corresponding
long-range correlations.}
\end{figure*}

The corresponding evolution of the Lagrangian radii and the interval
of negative specific heat are shown in Fig. \ref{kbkappa}.

In the dynamical friction scheme, energy dissipation depends, as in
the global dissipation scheme, on the absolute particle velocity.
Thus a strong dynamical friction extends the gravitational
collapse and the ``lifetime'' of the correlations. Yet, the observed
phase-space correlations are weaker, compared to those appearing
in simulations with global and local dissipation, respectively,
meaning that the reduced dissipation strength for fast particles,
accordingly to the dynamical friction scheme, destroys the
correlations.

\subsection{Effect of Short Distance Regularization}

So far the velocity correlations in dependence of the different
dissipative factors were studied.  Next the effect of different short
distance regularizations of the Newtonian potential, removing its
singularity, are checked.  That is, simulations with and without
short distance repulsive forces and with different dissipation
strengths are carried out. Energy is dissipated via the global
dissipation scheme.

\begin{figure*}
\centerline{
\psfig{file=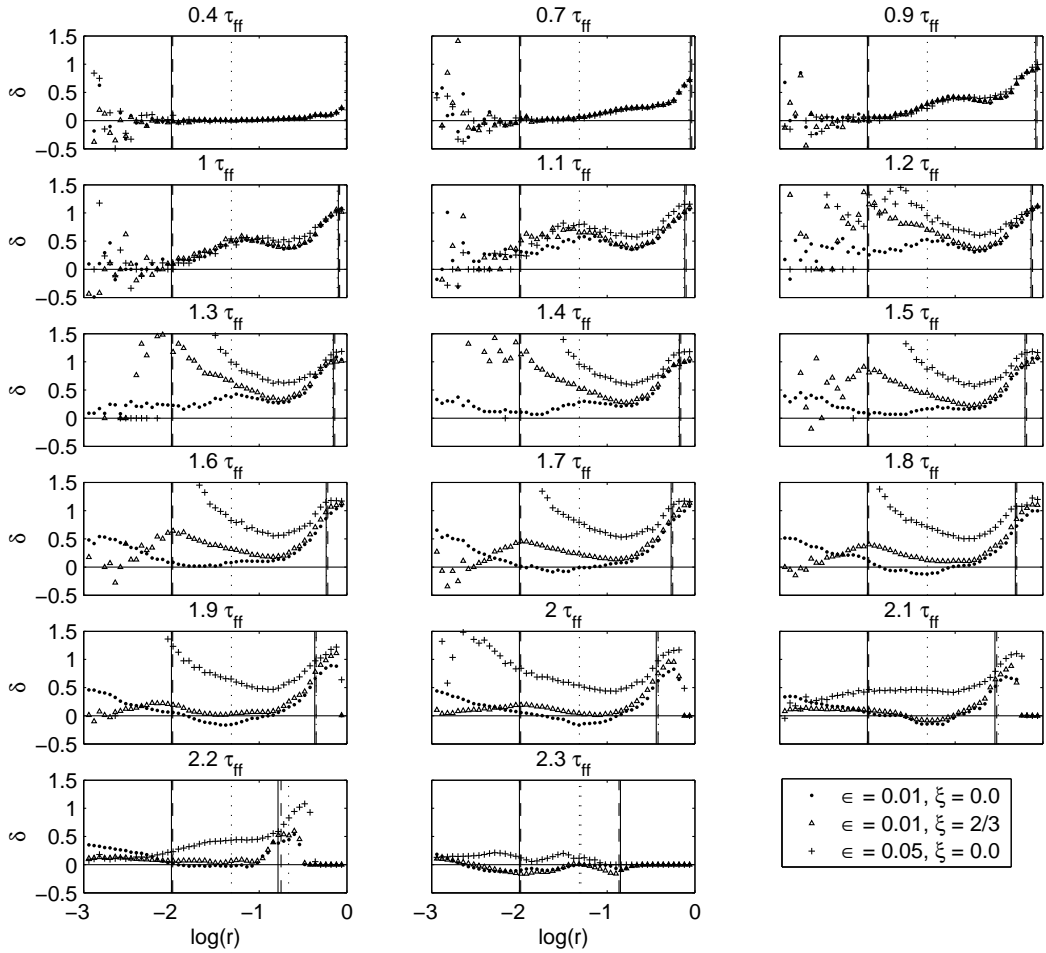,width=14cm}}
\caption{\label{kk6_1} Evolution of the velocity correlations 
  for three simulations with global dissipation
  scheme and different {\bf potential regularizations}. 
   That is, the softening 
  length and the form of the softening, respectively, 
  change from simulation to simulation. For $\xi=0.0$ the
  particle potential is a Plummer. $\xi=2/3$ means that there
  are short distance repulsive forces at work.
  The particle number is, N=160000.  
  The solid, the dashed and the dotted vertical
   lines indicate the scope of application of the 
   simulation with $(\epsilon=0.01,\xi=0.0)$, 
   $(\epsilon=0.01,\xi=2/3)$ and $(\epsilon=0.05,\xi=0.0)$,
   respectively. The
  evolution of the corresponding Lagrangian radii 
  are shown in Fig. \ref{kk6kappa}.}
\end{figure*}

\begin{figure*}
\centerline{
\psfig{file=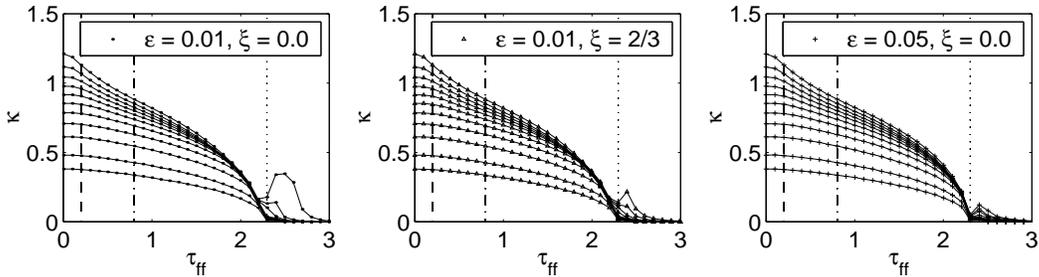,angle=90,width=14cm}}
\caption{\label{kk6kappa} Same as Fig. \ref{kbkappa} for three
simulations with different {\bf potential regularizations}, i.e., different
softening lengths and forms of the
softening potential, respectively.
The evolution of the corresponding
phase-space correlations are shown in Fig. \ref{kk6_1}.}
\end{figure*}

In Fig. \ref{kk6_1} the velocity correlations resulting from three
simulations with three different regularizations are compared.  The
regularizations are characterized by two parameters. Namely, the
softening length $\epsilon$ and the parameter $\xi$, that determines
the strength of the repulsive force. The softening length and $\xi$ of
the three simulations, compared in Fig. \ref{kk6_1}, are,
($\epsilon=0.01,\xi=0.0$), ($\epsilon=0.01,\xi=2/3)$ and
($\epsilon=0.05,\xi=0.0$), where $\xi=0.0$ means that a Plummer
potential is applied and $\xi>1/3$ means that short distance repulsive
forces are at work (see Sect. \ref{sece}). The dissipation strength is
for all three simulations the same, $\alpha=10$, and the collapsing
time is consequently the same as well (see Fig. \ref{kk6kappa}).

During the first $\tau_{\rm ff}$ long-range correlations resulting
from the three simulations are identical. Then the $\delta(r)$ starts
to separate. Indeed, the repulsive forces cause an increase of
$\delta$ at small scales, compared to the simulation with the same
softening length, but without repulsive forces. Yet, large scales
remain unaffected.

The index $\delta(r)$ resulting from the simulation with the large
softening length is larger compared to the other two simulations after
one $\tau_{\rm ff}$.  Also, the $\delta(r)$ curve is flatter after
that time.

The large difference between the simulation with the large softening
length and the corresponding simulation with the small softening
length is astonishing, because there is a clear deviation even at large
scales.  Well, the two simulations were carried out with the same
particle number, $N=160000$. Consequently, the volume filling factors are
different. Namely, $V_{\rm ff}=25$ and $V_{\rm ff}=0.16$, meaning that
for the large softening length, i.e., the large $V_{\rm ff}$, the mass
resolution is larger than the force resolution and vice versa for the
small softening length. Whereas a small volume filling factor
describes a granular phase, a large volume filling factor describes
rather a fluid phase. Thus the deviation of the velocity
correlations may mainly be due to the different volume filling factors
and not due to the different softening lengths.

In order to check this some complementary numerical experiments are
carried out, whose results are presented subsequently.

\subsection{Granular and Fluid Phase}

\begin{figure*}
\centerline{
\psfig{file=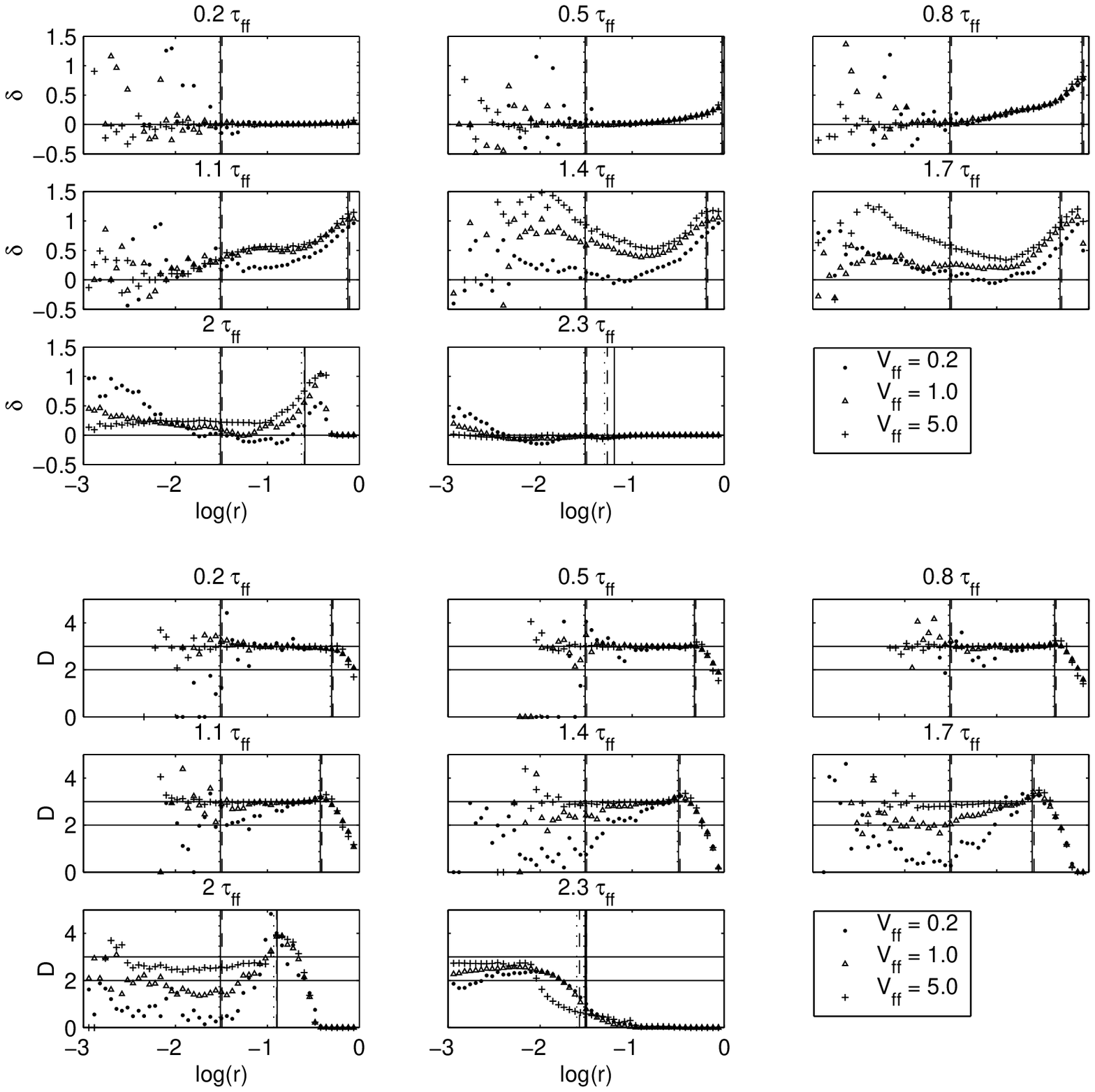,width=14cm}}
\caption{\label{mf_1} {\it Upper panels:} Evolution of the 
  index, $\delta(r)$, of the velocity-dispersion-size relation,
  $\sigma\propto r^{\delta(r)}$, 
  resulting from three
  simulations with different {\bf volume filling factor} that describe a
  granular phase, $V_{\rm ff}=0.2$, a fluid phase, $V_{\rm ff}=5.0$, and a
  intermediate state, $V_{\rm ff}=1.0$. The particle numbers are 
  $N=6400$, $N=160000$ and $N=32000$, respectively. The softening
  length is $\epsilon=0.031$. The solid,
  the dashed and the dotted vertical lines indicate the scope of
  application of the simulation with $V_{\rm ff}=0.2$, $V_{\rm ff}=1.0$ and
  $V_{\rm ff}=5.0$, respectively. The scope is give by,
  $\epsilon<r<2R_{90}$. The corresponding Lagrangian radii are
  presented in Fig. \ref{mfkappa}. {\it Lower panels:} The evolution
  of the index, $D(r)$, of the mass-size relation, $M\propto r^{D(r)}$. 
  The solid,
  the dashed and the dotted vertical lines indicate the scope of
  application of the simulation with $V_{\rm ff}=0.2$, $V_{\rm ff}=1.0$ and
  $V_{\rm ff}=5.0$, respectively. The scope is give by,
  $\epsilon<r<R_{90}/2$.}
\end{figure*}

The upper panels in Fig. \ref{mf_1} show the velocity-dispersion-size
relation, that develop in simulations with different volume filling
factors, but equal softening length, $\epsilon=0.031$. The particle
numbers are, $N=6400$, $N=32000$ and $N=160000$.  Thus, the volume
filling factors are, $V_{\rm ff}=0.2$, $V_{\rm ff}=1.0$ and $V_{\rm
  ff}=5.0$, respectively. Strongest phase-space correlations appear in
the simulation with the highest mass resolution, i.e., for the ``fluid
phase'', and weakest correlations appear in the ``granular phase''.
The flattest $\delta(r)$ curve results from the simulation in which
force and mass resolution are equal.

This result suggests that the volume filling factor $V_{\rm ff}$ (and
not the softening length) is the crucial parameter determining the
correlation strength in simulations with equal dissipation strength,
as long as a substantial part of the system forces stem from
interactions of particles with relative distances larger than the
softening length.

The lower panels in Fig. \ref{mf_1} show, for the same simulations,
the evolution of the index, $D(r)$, of the mass-size relation,
$M\propto r^{D(r)}$. The deviation from homogeneity is strongest for
the ``granular phase'' and weakest for the ``fluid phase''. Thus the
order of the correlation strength in space is inverse to the order of
the correlation strength in phase-space. Such an inverse order is
expected in self-gravitating system with $|U|\propto T$, where $U$ is
the potential energy and $T$ is the kinetic energy (Pfenniger \&
Combes \cite{Pfenniger94}; Pfenniger \cite{Pfenniger96}; Combes
\cite{Combes99}).

\begin{figure*}
\centerline{
\psfig{file=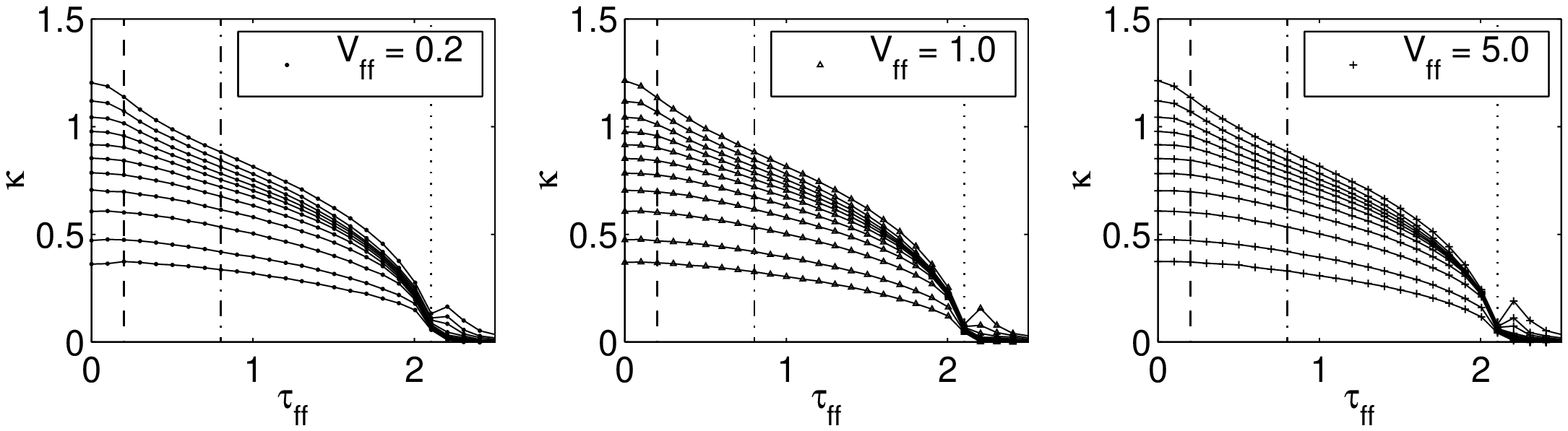,angle=90,width=14cm}}
\caption{\label{mfkappa} Same as Fig. \ref{kbkappa} for three
simulations with different {\bf volume filling factor}.
Fig. \ref{mf_1} shows the evolution of the corresponding 
long-range correlations.}
\end{figure*}

The corresponding Lagrangian radii and the interval of negative
specific heat are shown in Fig. \ref{mfkappa}.

\subsection{Initial Noise}

Assuming a constant softening length, different volume filling
factors, $V_{\rm ff}$, mean different particle numbers, $N$, that
introduce different Poisson noise, $\sim \sqrt{N}$.  Then, the
statistical roughness of the initial uniform Poisson particle
distribution decreases like $\sim 1/\sqrt{N}$.

Thus the initial roughness of the latter simulations (see Fig. \ref{mf_1})
differ from each other by a factor $\sim 2.2$ and one might suppose 
that the different correlation strengths in these simulations are
the result of the different statistical, initial roughness. 

\begin{figure*}
\centerline{
\psfig{file=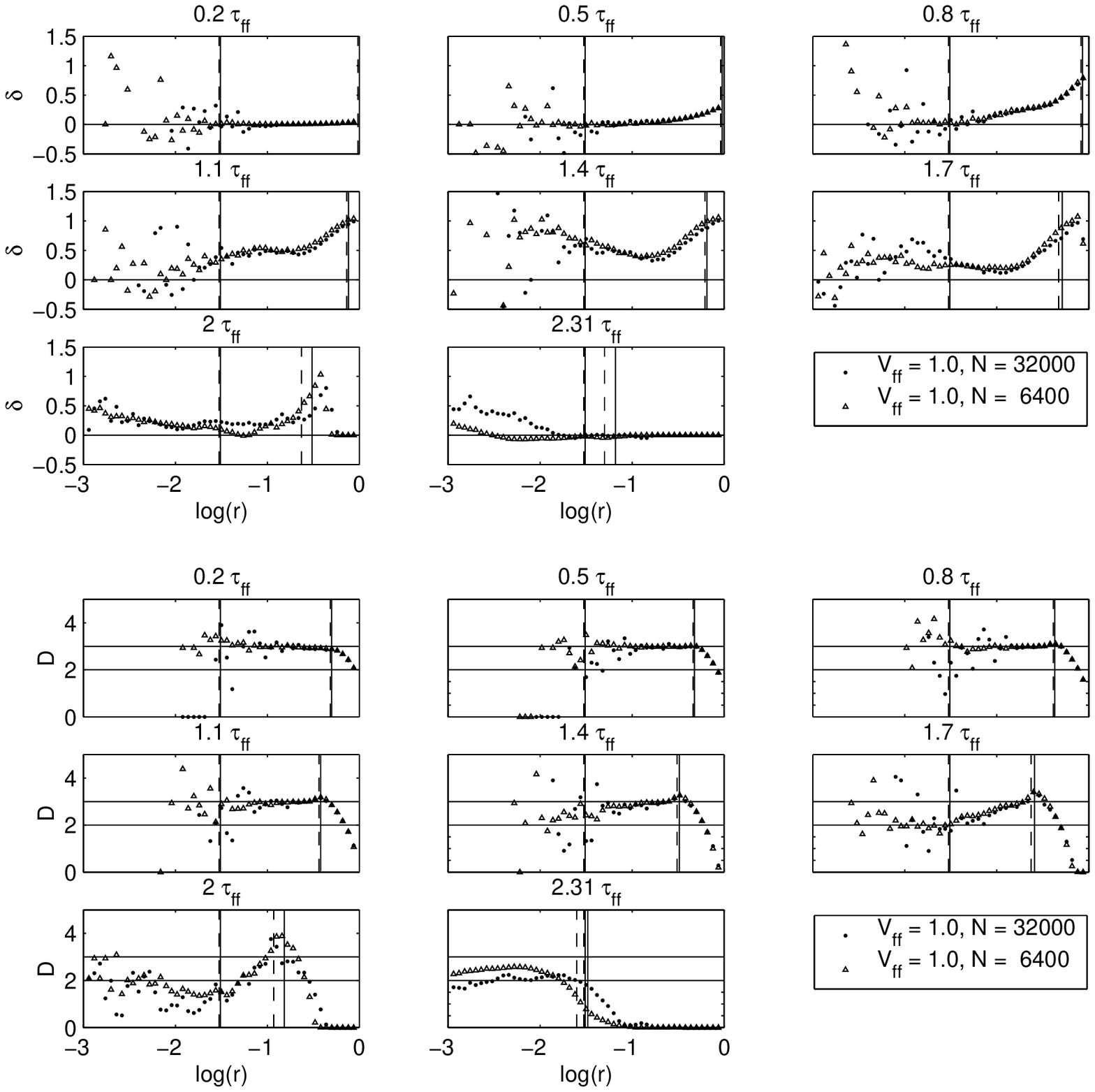,width=14cm}}
\caption{\label{mf2_1} {\it Upper panels:} Evolution of the
phase-space correlations resulting, resulting from two
simulations with identical volume filling factor, 
$V_{\rm ff}=1.0$, but unequal statistical, {\bf initial roughness}, 
$\sim 1/\sqrt{N}$. The particle number and the softening length of the
first simulation (dots) are $N=32000$ and $\epsilon=0.031$, 
respectively. The second simulation (triangles) is carried out with
$N=6400$ and $\epsilon=0.054$. The solid and the dashed
lines indicate the scope of application given by
$\epsilon<r<2R_{90}$. The evolution of the corresponding
Lagrangian radii is shown in Fig. \ref{mf2kappa}.
{\it Lower panels:} The evolution of the spatial correlations.
The scopes of application, depicted by the vertical lines, is,
$\epsilon<r<R_{90}/2$.}

\end{figure*}

\begin{figure*}
\centerline{
\psfig{file=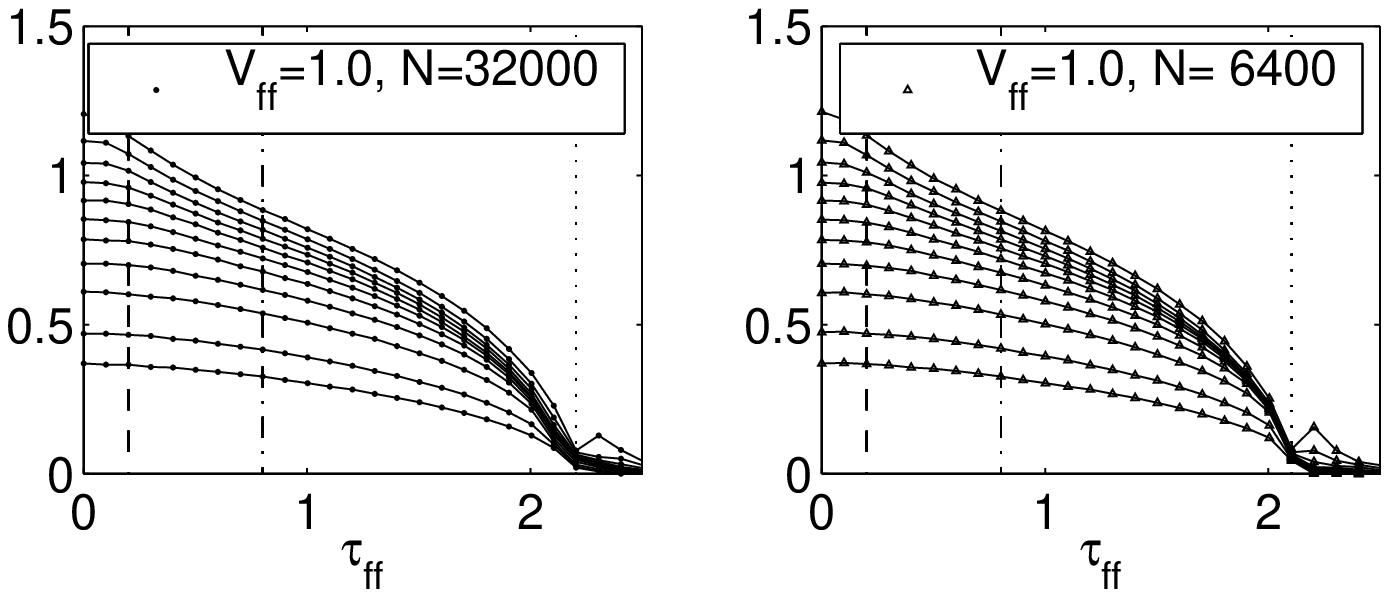,angle=90,width=9.3cm}}
\caption{\label{mf2kappa} Same as Fig. \ref{kbkappa} for two
simulations with identical volume filling factor and 
unequal {\bf initial roughness}, $\sim 1/\sqrt{N}$.
The evolution of the corresponding
long-range correlations is shown in Fig. \ref{mf2_1}}
\end{figure*}

In order to check this possibility, long-range correlations are
compared that result from two simulations with $V_{\rm ff}=1.0$ and
with statistical roughness that differs by a factor $\sim 2.2$. The
results are presented in Fig. \ref{mf2_1} and the corresponding
Lagrangian radii are shown in Fig. \ref{mf2kappa}. The simulation with
$N=32000$ was already presented in Fig. \ref{mf_1}. It is now compared
with a simulation with stronger initial roughness.

Despite the different initial roughness, long-range correlations in
phase-space are almost identical for the two simulations.

As regards
fragmentation, additionally to the parameter of the
mass-size relation, the mass distribution in space were compared. 
We find actually a stronger fragmentation for small particle numbers.
Yet, the $V_{\rm ff}$ seems to be the crucial parameter for the
fragmentation strength in Fig. \ref{mf_1}.  

Consequently, the different correlation strengths in Fig. \ref{mf_1}
can not be accounted for by Poisson noise, but are mainly the result
of the different volume filling factors, or more precisely, due to
the different ratio of force and mass resolution. Thus
dark matter clustering in high-resolution cosmological $N$-body
simulations in which the force resolution is typically an order
of magnitude smaller than the mass resolution may be too strong
compared with the physics of the system 
(Hamana et al. \cite{Hamana01}).  

\subsection{Free Fall of Cold Dissipationless Systems}

Here, long-range correlations are discussed that develop in cold,
gravitationally unstable systems without energy dissipation. The
correlations strength appearing in such dissipationless systems depends
on the initial ratio between kinetic and potential energy, $a=T/|U|$,
i.e., on the number of thermal Jeans masses given through $M/M_{\rm
  J}=2a^{-3/2}$.

\begin{figure*}
\centerline{
\psfig{file=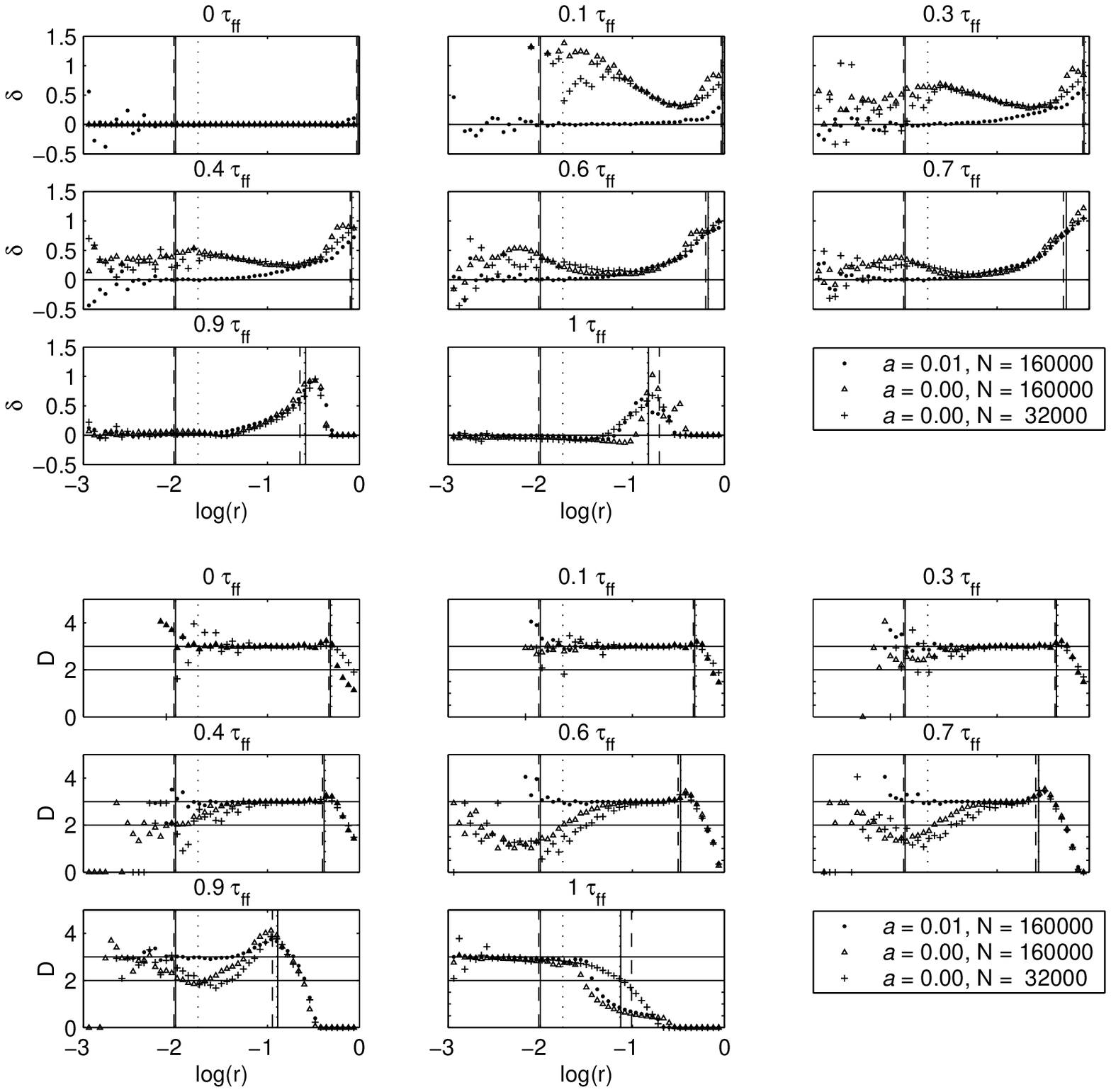,width=14cm}}
\caption{\label{ff_1} {\it Upper panels:} Phase-space correlations
that develop during the {\bf free fall} of three cold, 
dissipationless systems. All systems have the same volume
filling factor, $V_{\rm ff}=0.016$. The parameter $a$ indicates 
the ratio between kinetic and potential energy.
The solid, the dashed and the dotted 
lines indicate the scope of application,
$\epsilon<r<2R_{90}$, of the 
simulation with $(a=0.01,N=160000)$, 
$(a=0.0,N=160000)$ and $(a=0.0,N=32000)$,
respectively. The evolution of the corresponding
Lagrangian radii is shown in Fig. \ref{ffkappa}.
{\it Lower panels:} Evolution of the spatial correlations.
The scopes of application, depicted by the vertical lines, is,
$\epsilon<r<R_{90}/2$.}
\end{figure*}

This is shown in Fig.  \ref{ff_1}. For $a=0.01$ the index $\delta$ of
the velocity-dispersion-size relation remains zero at small scales
during the whole free fall, whereas for $a=0.0$ the index becomes,
$\delta>0$, over the whole dynamical range, during a period of $\sim 0.7
\tau_{\rm ff}$. That is, 350 $M_{\rm J}$ are not sufficient to develop small
scale phase-space correlations in a dissipationless system. 

In order to show the dependence of the result on the initial
Poisson noise, the absolutely cold simulation with $N=160000$ is
compared with a $N=32000$-body simulation.  As above, it follows that
the initial roughness of the two systems differs by a factor $\sim
2.2$.  

Fig. \ref{ff_1} shows that the non-equilibrium structures resulting
form the simulations with equal $a$ but unequal particle number differ
from each other during the whole free fall. Indeed, during the first
half of the free fall time the two initial conditions produce velocity
correlations that differ on small and large scales.  After
$t=0.5\;\tau_{\rm ff}$ the differences approximately disappear.  However,
the behavior of the spatial correlations (see lower panels of Fig.
\ref{ff_1}) is inverse, meaning that differing spatial correlations
appear after $0.5\;\tau_{\rm ff}$ and persist for the rest of the free
fall.

These results suggest that non-equilibrium structures appearing in 
cold systems or in systems with very effective energy dissipation 
depend more strongly on initial noise than those appearing in warm
systems with less effective dissipation (see also Fig. \ref{mf2_1}).

\begin{figure*}
\centerline{
\psfig{file=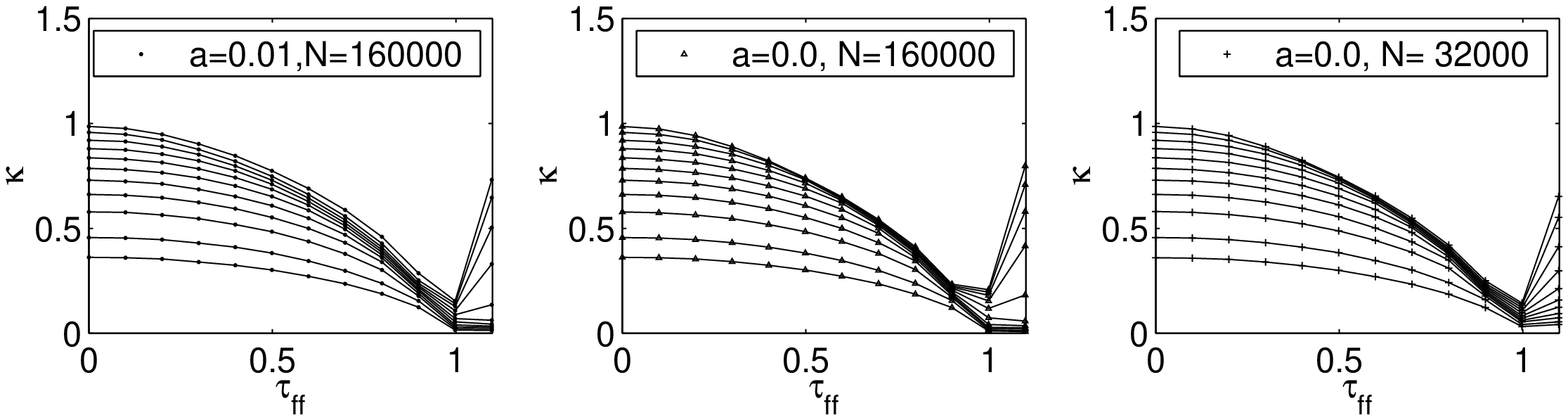,angle=90,width=14cm}}
\caption{\label{ffkappa} Same as Fig. \ref{kbkappa} for the 
{\bf free fall} of three
cold, dissipationless systems with identical volume 
filling factor but unequal $a$ and initial roughness, 
respectively. The evolution of the corresponding
long-range correlations is shown in Fig. \ref{ff_1}.}
\end{figure*}

The evolution of the Lagrangian radii during the free fall of the
cold, dissipationless systems are shown in Fig. \ref{ffkappa}. Because
these systems are adiabatic and self-gravitating during the whole
simulation, the corresponding intervals are not plotted in the figure.

\subsection{Discussion}

Long-range spatial and phase-space correlations appear naturally
during the collapsing phase transition in the interval of negative
specific heat if the energy dissipation time is, $\tau_{\rm
  dis}\la\tau_{\rm ff}$, so that the time-scale of correlation
growth is smaller than the time-scale of chaotic mixing which
is, $\sim 1/\lambda\propto\tau_{\rm ff}$, where $\lambda$
is the maximum Liapunov exponent (Miller \cite{Miller94}).

Actually, the details of the long-range correlations depend on the
applied dissipation scheme, but there are also some generic
properties. That is, phase space correlations start to grow at large
scales, whereas spatial correlations seem to grow bottom-up.
Moreover, there is an upper limit for the index of the
velocity-dispersion-size relation within the dynamical range, namely,
$\delta\approx 1$.

Besides, the dissipation strength, and initial conditions, the volume
filling factor $V_{\rm ff}$ is a crucial parameter for the correlation
strength. That is, phase-space correlations are strong in the fluid
limit and weak for a granular phase. The behavior of the spatial
correlations is exactly the other way around.

The softening length does not affect correlations within the dynamical
range. Yet, sub-resolution repulsive forces affect correlations on
small scales above the resolution scale. Of course, this does not hold
for the onset of the correlation growth, but only after sufficient
particles have attained sub-resolution distances.

The above considerations suggest that the found correlations
are physically relevant and not a numerical artifact. 

The long-range spatial and phase-space correlations appearing during
the collapsing transition are qualitatively similar to the mass-size
and the velocity-dispersion-size relation observed in the ISM (e.g.
Blitz \& Williams \cite{Blitz99}; Chappell \& Scalo \cite{Chappell01};
Fuller \& Myers \cite{Fuller92}), showing that in the models gravity
alone can account for ISM-like correlations.

Furthermore, the time-scale of the correlation lifetime is the free
fall time, $\tau_{\rm ff}$, which is consistent with a dynamical
scenario in which ISM-structures are highly transient
(V\'azquez-Semadeni \cite{Vazquez02}; Larson \cite{Larson01}; Klessen
et al. \cite{Klessen00}), which is related with rapid star formation
and short molecular cloud lifetimes (Elmegreen \cite{Elmegreen00}),
that is, the corresponding time-scales are about an order of magnitude
smaller than in the classical Blitz \& Shu (\cite{Blitz80}) picture.

\section{Results: III. Permanent Energy-Flow}

\subsection{Nonequilibrium Structures in Systems Subject 
to an Energy Flow}
 
The numerical experiments presented above showed that dissipative
self-gravitating systems fragment and establish long-range
correlations outside of equilibrium in the interval of negative
specific heat. These transient correlations persist for
$1-2\;\tau_{\rm ff}$ times.

Here we check if self-gravitating systems can establish persistent
long-range correlations if they are maintained continuously outside of
equilibrium by a permanent energy-flow. That is, the dissipated energy
is continuously replenished by time-dependent potential perturbations.

Both simulations of granular and fluid phases are carried out. A
typical parameter set, describing a granular system is, $N=10000$,
$\epsilon=0.0046$, $V_{\rm ff}=0.001$ and $N=160000$,
$\epsilon=0.0315$, $V_{\rm ff}=5$ are here typical for a fluid phase.
Yet, the parameters are not fixed and the effect on the evolution is
studied when parameters change. Parameters, controlling energy-flow
and interaction potential, as well as their ranges are indicated in
Table \ref{tab1}.

The applied potential perturbations imitate massive objects passing in
the vicinity on time-scales, $\tau_{\rm pert}\la\tau_{\rm
  dyn}$. The perturbations induce primarily ordered particle motions.
Then gravitational interactions lead to a conversion of the bulk
kinetic energy to random thermal motion. The energy injection due to
such a forcing scheme can be quite regular until a plateau is reached.

\begin{figure*}
\centerline{
\psfig{file=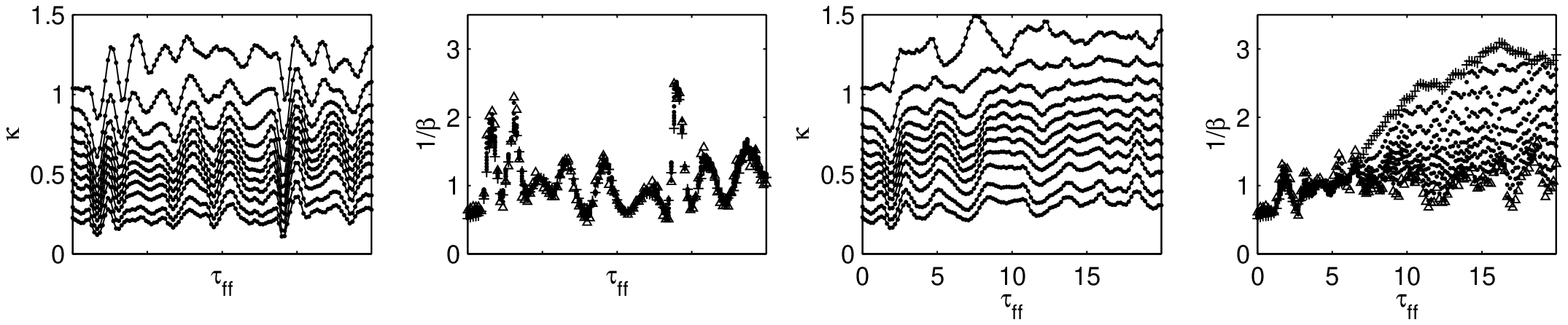,angle=90,width=\hsize}}
\caption{\label{beta} {\it The two left panels:} Evolution of the 
  Lagrangian radii (left) and the Lagrangian 
  dimensionless temperature (middle left) 
  of a system subjected to an energy-flow and with global 
  dissipation scheme. The curves depict certain mass-fractions,
  and its temperatures, respectively.
  The mass fractions, contained in spheres, centered at the center of
  mass, are:
  $\Delta M/M=\{5\%, 10\%, 20\%, \ldots ,80\%, 90\%, 98\%\}$. 
  The triangles (left panel) indicate the temperature of a mass 
  fraction of $5\%$ and the crosses corresponds to a mass fraction 
  of $98\%$. The simulation is carried out with $N=10000$ and 
  $\epsilon=0.0046$. Consequently, $V_{\rm ff}=0.001$, meaning that
  the system is highly granular. The dynamical range is, 2.3 dex. 
  {\it The two right panels:} Same as above for a system subjected
  to an energy-flow and with a local dissipation scheme.}
  \vspace{1.5ex}
\end{figure*}

Energy injection prevents a system from collapsing and maintains an
approximately statistically steady state during $\sim5-15\;\tau_{\rm
  dyn}$ when energy dissipation is balanced appropriately by large
scale potential perturbations. These states do not feature any
persistent long-range correlations. Yet, they develop a temperature
structure that is characteristic for the applied dissipation scheme.
This is shown in Fig. \ref{beta}, where the evolution of two granular
systems subjected to an energy-flow is presented.  One system
dissipates its energy by a global dissipation scheme (top), the other
by a local dissipation scheme (bottom).

The system with the global dissipation scheme is nearly thermalized
during almost the whole simulation, whereas the local dissipation
scheme leads to a permanent positive temperature gradient, that is
inverse compared to stars and resembles those of the ISM where dense,
cool mass condensation are embedded in hotter shells.

\begin{figure*}
\centerline{
\psfig{file=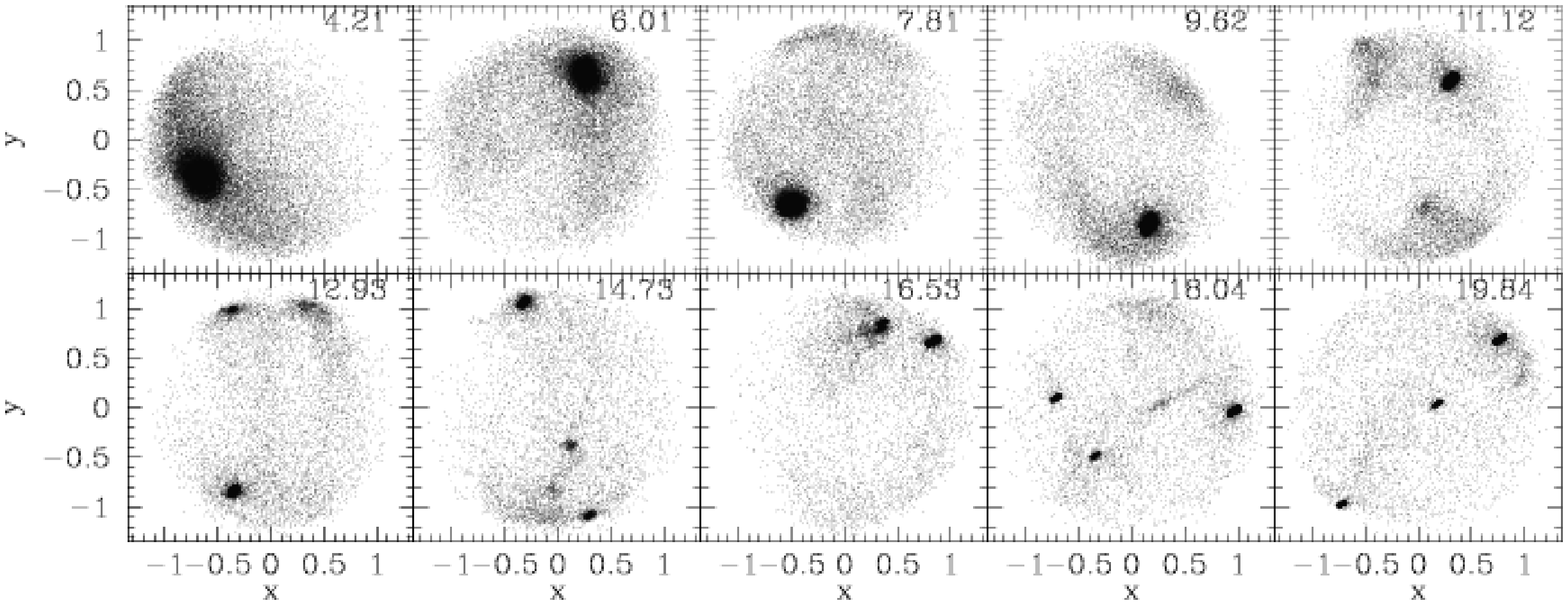,angle=-90,width=18cm}}
\caption{\label{eflow} Mass distribution of a system subjected to 
an energy-flow. Shown is the projection of the
particle positions onto the $xy$-plane.
The particle number is, $N=32000$, and the 
volume filling factor is, $V_{\rm ff}=1$. Energy is dissipated
with a local dissipation scheme, $\beta=150$, and short 
distance repulsive forces are at work, $\xi=2/3$.  
The evolution of the corresponding phase-space correlations
is shown in Fig. \ref{eflowcor}. The time is indicated in each panel
in units of the free fall time, $\tau_{\rm ff}$.}
\end{figure*}

\begin{figure*}
\centerline{
\psfig{file=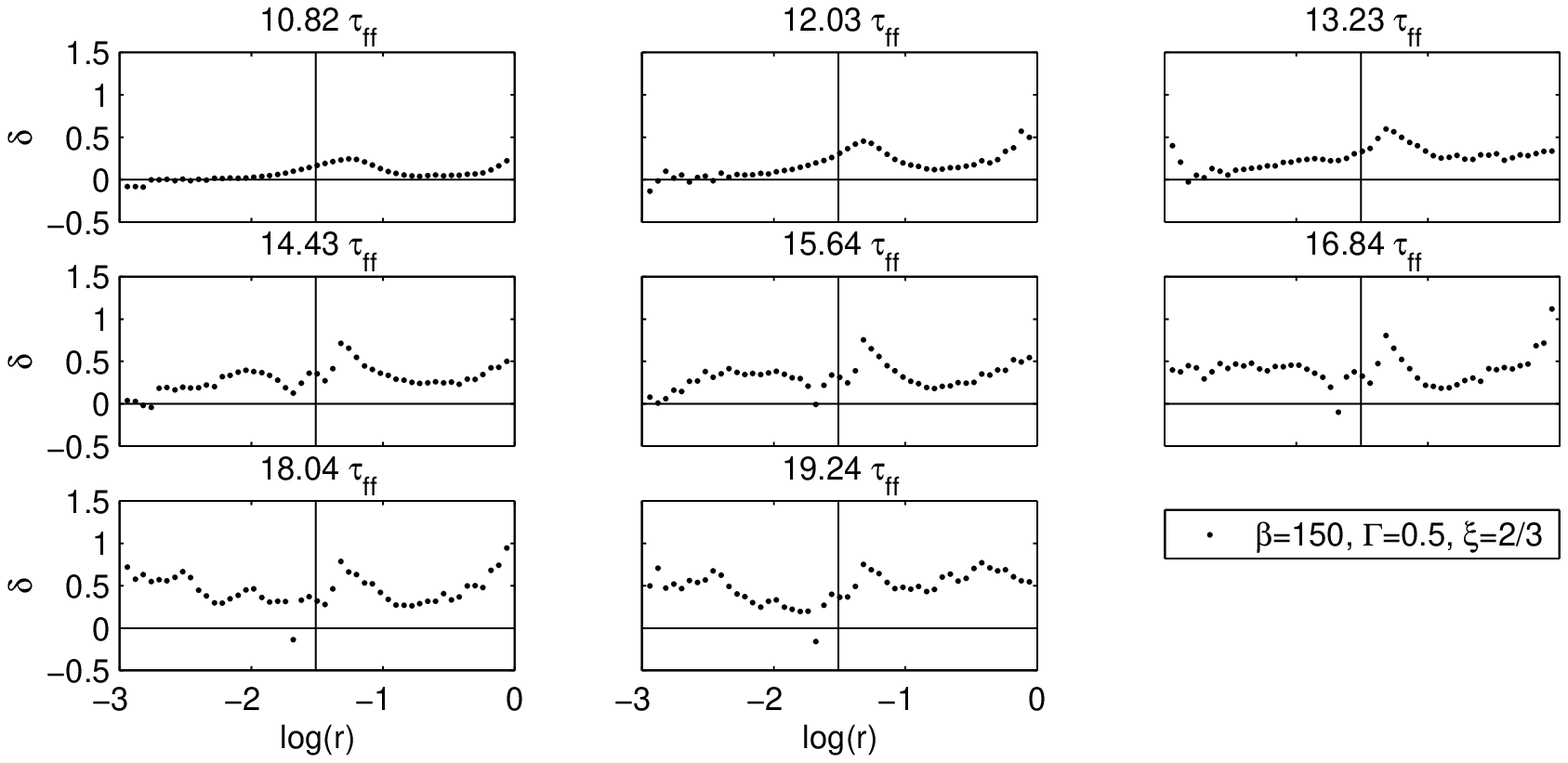,width=14cm}}
\caption{\label{eflowcor} The evolution of the phase-space
  correlations in a system subjected to an energy-flow and with short
  distance repulsive forces (see Fig. \ref{eflow}).  During the first
  $\sim 8\;\tau_{\rm ff}$ the index of the velocity-dispersion-size
  relation is, $\delta=0$, over the whole studied range, then
  correlations start to grow at largest scales.  Finally, the index
  of the velocity-dispersion-size relation is, $\delta>0$, over the 
  whole dynamical range. In systems subjected to energy-flows, 
  such correlations that extend over the whole dynamical range appear 
  only if a local dissipation scheme is applied.}
\end{figure*}

If dissipation dominates energy injection the system undergoes in
general a mono-collapse, that is, a collapsed structure is formed in
which a part of the system mass is concentrated in a single dense
core and the rest is distributed in a diffuse halo. However, 
systems with a rather fluid phase may develop several dense cores moving
in a diffuse halo, when they are subjected to an appropriate energy
flow. In the course of time the number of clumps varies,
but a non-mono-clump structure persists for some $~\tau_{\rm ff}$ (see
Fig. \ref{eflow}).  These systems may even develop persistent
phase-space correlations (see Fig. \ref{eflowcor}).

However, the clumps result not from a hierarchical fragmentation
process, but they are formed sequentially on the free fall time scale.
Furthermore, the clumps are so dense that their evolution is strongly
influenced by the applied regularization. That is, the evolution over
several $\tau_{\rm ff}$ depend on the numerical model that does not
represent accurately small-scale physics.

In order to impede gravitational runaway that may hinder the formation
of complex non-equilibrium structures within the given dynamical range,
different measures are taken, such as short distance repulsion and the
application of the dynamical friction scheme, where the friction force
$F\to 0$ for $v\to\infty$.

Yet, despite these measure, we only find either nearly homogeneous
structures in systems where energy injection prevails, or systems
dominated by unphysical clumps in case of prevailing energy dissipation.

Up to now, mainly the effect of several different dissipation schemes
were discussed. However, the effect of a modified forcing scheme is
also checked.  That is, a power-law forcing scheme is applied, which
injects energy at different frequencies. This forcing scheme is a
modification of those presented in Sect.~\ref{heating} and reads,
$\Phi_{\rm pert}(\omega)\propto\omega^\nu$, where $\nu=-4,\ldots,4$.
Yet, also such a forcing scheme can not induce a phase transition to
complex non-equilibrium structures and the resulting mass distribution
corresponds to the those described above.

\subsection{Discussion}

Actually the ``clumpy'' structure in Fig. \ref{eflow} and the
corresponding phase-space correlations shown in Fig. \ref{eflowcor} do
not represent real physics, nevertheless, they show that it is
principally possible to maintain spatial non-equilibrium structures
and long-range phase-space correlations in a perturbed, dissipative,
self-gravitating system over several dynamical times. Thus, it can not
be excluded that in the future, with a better representation of
microscopic physics and forcing mechanisms at work in the ISM, models
including self-gravity may produce complex nonhomogeneous structures
in a statistical equilibrium, i.e., persistent patterns formed by
transient structures.

However, at present models of dissipative, self-gravitating systems,
cannot produce such structures on the scale of Giant Molecular Clouds
(e.g. Semelin \& Combes \cite{Semelin00}, Klessen et al. \cite{Klessen00},
Huber \& Pfenniger \cite{Huber01a}).

On larger scales, gravity gives rise to persistent non-equilibrium
structures in cosmological and shearing box simulations. A common
denominator of these models is that their time dependent boundary
conditions are given by a scale-free spatial flow counteracting
gravity. Let us discuss this more precisely.

In cosmological and shearing box models time-dependent boundary
conditions create relative particle velocities that are inverse to
gravitational acceleration and increase with particle distance,
$v\propto r$. In the shearing box model, the relative azimuthal
particle velocity due to the shear flow is, $v_{\theta}\propto r_c$,
where $r_c$ is the radial particle distance in cylinder coordinates.
In cosmological models, the relative particle velocity induced by the
Hubble flow is, $v_r \propto r$, where $r$ is the relative particle
distance in Cartesian coordinates. 

These relations and consequently the corresponding flows are
scale-free. The fact that the shear flow affects only the azimuthal
velocity component may then account for the characteristic spiral arm
like structures found in shearing box experiments, differing from
those found in cosmological models, where the isotropic Hubble flow
gives rise to in average isotropic non-equilibrium structures.

The models studied in this paper are not subject to a scale-free
spatial flow counteracting gravity and persistent long-range
correlations of astrophysical relevance do not appear. This may
suggest that in situations where gravitational runaway is allowed,
matter that has passed through a collapsing transition, has to be
replenished at large scales in order to attain a statistical
equilibrium state of transient fragmentation.

\section{Conclusions}

First, equilibrium states of $N$-body models were compared with 
analytical models. Subsequently, the findings resulting from
this comparison are summarized:

\begin{itemize} 

\item On the one hand, equilibrium properties of $N$-body models agree with
predictions made by analytical models. An example is the energy
interval of negative specific heat. One the other hand, discrepancies
were found, such as the way the collapsing phase transition,
separating a high-energy homogeneous phase from a low-energy collapsed
phase, develops in the interval of negative specific heat. These
discrepancies suggest: 1.) Small scale physics becomes relevant
for the system evolution when the growth of singularities triggered
by gravitational instabilities is allowed. 
2.) Analytical models based on the Gibbs-Boltzmann entropy are not 
strictly applicable to non-extensive self-gravitating systems. 
Yet, not all of the equilibrium properties found by maximizing
the Gibbs-Boltzmann entropy are expected to change if a fully
consistent, generalized thermostatistical theory  is applied.   

\end{itemize}

\noindent Second, the collapsing transition was studied in systems with strong
dissipation. The findings are:

\begin{itemize} 
  
\item Dissipative self-gravitating systems develop outside of
  equilibrium, in the interval of negative specific heat, transient
  long-range correlations. That is, fragmentation and nonequilibrium
  velocity-dispersion-size relations, with striking resemblance to those
  observed in the ISM, appear during the collapsing transition, when
  the dissipation time is shorter than the dynamical time. This
  suggests that nonequilibrium structures in self-gravitating
  interstellar gas are {\it dynamical and highly transient}.
  
\item Besides the dissipation strength and the initial noise, the
  granularity turns out to be a crucial parameter for the strength of
  the resulting long-range correlations, substantiating the
    importance of a coherent mass and force resolution. That is,
  phase-space correlations are stronger in the fluid limit than in a
  granular phase. The opposite holds for spatial correlations.  This
  substantiates The inverse behavior of fragmentation strength and
  phase-space correlation strength is found in all simulations and is
  typical for self-gravitating systems.

\end{itemize}

\noindent Finally, systems subject to a permanent energy-flow were
studied. We find:

\begin{itemize}
  
\item Typically driven dissipative systems evolve to a high-energy
  homogeneous phase or undergo a mono-collapse. Yet, model systems
  with a local energy dissipation can develop persistent phase-space
  correlations, but a persistent, hierarchical fragmented structure is
  not observed. This suggests, that matter that has passed through a
  collapsing transition has to be replenished at large scales in order
  to maintain a hierarchical structure at molecular cloud scales.

\end{itemize}

\begin{acknowledgements}

This work has been supported by the Swiss National Science Foundation.

\end{acknowledgements}

\end{document}